\documentclass[12pt]{iopart}
\usepackage{iopams}  
\expandafter\let\csname equation*\endcsname\relax
\expandafter\let\csname endequation*\endcsname\relax
\usepackage{amsmath}
\usepackage{amsfonts}
\usepackage{mathtools}
\usepackage{graphicx}
\usepackage{leftidx}
\usepackage{ulem}
\usepackage{xcolor, soul}

\usepackage[colorlinks=true, allcolors=blue]{hyperref}

\newcommand\underrel[3][]{\mathrel{\mathop{#3}\limits_{%
      \ifx c#1\relax\mathclap{#2}\else#2\fi}}}

\begin{document}

\title{The OU$^2$ process: Characterising dissipative confinement in noisy traps}

\author{Luca Cocconi\textsuperscript{1}, Henry Alston\textsuperscript{2}, Jacopo Romano\textsuperscript{1} and Thibault Bertrand\textsuperscript{2}}
\address{$^1$ Max Planck Institute for Dynamics and Self-Organization (MPIDS), 37077 G{\"o}ttingen, Germany\\
$^2$ Department of Mathematics, Imperial College London, South Kensington, London SW7 2AZ, United Kingdom}

\ead{luca.cocconi@ds.mpg.de}
\vspace{10pt}
\begin{indented}
\item[]\today
\end{indented}

\begin{abstract}
The Ornstein-Uhlenbeck (OU) process describes the dynamics of Brownian particles in a confining harmonic potential, thereby constituting the paradigmatic model of overdamped, mean-reverting Langevin dynamics. Despite its widespread applicability, this model falls short when describing physical systems where the confining potential is itself subjected to stochastic fluctuations. However, such stochastic fluctuations generically emerge in numerous situations, including in the context of colloidal manipulation by optical tweezers, leading to inherently out-of-equilibrium trapped dynamics. To explore the consequences of stochasticity at this level, we introduce a natural extension of the OU process, in which the stiffness of the harmonic potential is itself subjected to OU-like fluctuations. We call this model the OU$^2$ process. We examine its statistical, dynamic, and thermodynamic properties through a combination of analytical and numerical methods. Importantly, we show that the probability density for the particle position presents power-law tails, in contrast to the Gaussian decay of the standard OU process. In turn, this causes the trapping behavior, extreme value statistics, first passage statistics, and entropy production of the OU$^2$ process to differ qualitatively from their standard OU counterpart. Due to the wide applicability of the standard OU process and of the proposed OU$^2$ generalisation, our study sheds light on the peculiar properties of stochastic dynamics in random potentials and lays the foundation for the refined analysis of the dynamics and thermodynamics of numerous experimental systems.
\end{abstract}

%\noindent{\it Keywords}: fluctuating potentials, first passage statistics, 


%\maketitle


% For a table of contents, uncomment the next line
%\tableofcontents

%%%%%%%%%%%%%
% Section: Introduction %
%%%%%%%%%%%%%
\section{Introduction}
\label{sec:introduction}

The Ornstein-Uhlenbeck (OU) process is a continuous-time Gaussian stochastic process with linear mean-reverting properties first introduced to describe the fluctuating velocity of a Brownian particle immersed in a fluid \cite{Uhlenbeck1930,Einstein1905}. It has found over the years countless applications across various subfields of physics (as well as other disciplines), where it plays a similarly paradigmatic role as that of the harmonic oscillator in classical and quantum mechanics. It can be understood as the overdamped limit of any Langevin dynamics exploring the local neighborhood of a differentiable minimum of an arbitrary potential landscape. Its Langevin equation of motion generically reads
\begin{equation}\label{eq:original_OU}
    \frac{d x(t)}{dt} = - \bar{k} x(t) + \sqrt{2 D_x} \zeta(t)\,.
\end{equation}
where, depending on the context, $\bar{k}>0$ might be interpreted as a friction coefficient or potential stiffness coefficient, while $D_x$ denotes the diffusivity and $\zeta(t)$ is a delta-correlated zero mean and unit variance white noise. Key results including steady-state probability density function, formal solution and Green's function for the standard OU process are reviewed for completeness in \ref{sec:keyOU}. To give but one example of its wide applicability for instance in spatially extended settings, a lattice of elastically coupled OU processes formally defines the Gaussian free field around which the perturbative expansion of non-conserved dynamical field theories is typically constructed \cite{LeBellac1991,Hohenberg1977}. 

While originally modelling the frictional contribution to Brownian motion, the first term in the right-hand side of Eq.~\eqref{eq:original_OU} is often interpreted in the case of overdamped dynamics as a restoring force resulting from an effective harmonic potential $V(x) = \bar{k}x^2/2$ acting on the coordinate $x$. This is the case, for instance, in many physical models of micro-particle manipulation by optical tweezers \cite{Martinez2017,Saha2022,Saha2021,Saha2023,TalFriedman2020}, where force gradients are established through the inhomogeneous electric field within a highly focused laser beam \cite{Ashkin1986,Dufresne2001,Bustamante2021,Pesce2020}. However, the nature of the potential $V(x)$ might even be more abstract, as exemplified by models of mean reverting portfolios in finance \cite{Chiang1995} or continuous trait evolution in ecology \cite{Bartoszek2017}. 

In all such cases, it is reasonable to expect that the underlying processes governing the potential are themselves subject to some degree of stochasticity, implying that $V(x,t)$ may itself be a stochastic process \cite{Alston2022,Cocconi2023}. A case in point is that of optical tweezers controlled by real laser systems, which are characterised by small fluctuations in power output around its mean \cite{Bustamante2021,Pesce2020}; these fluctuations in power lead in turn to fluctuations in the stiffness of the potential experienced by the dielectric particle.

\begin{figure}[t!]
    \centering
    \includegraphics[width=0.9\textwidth]{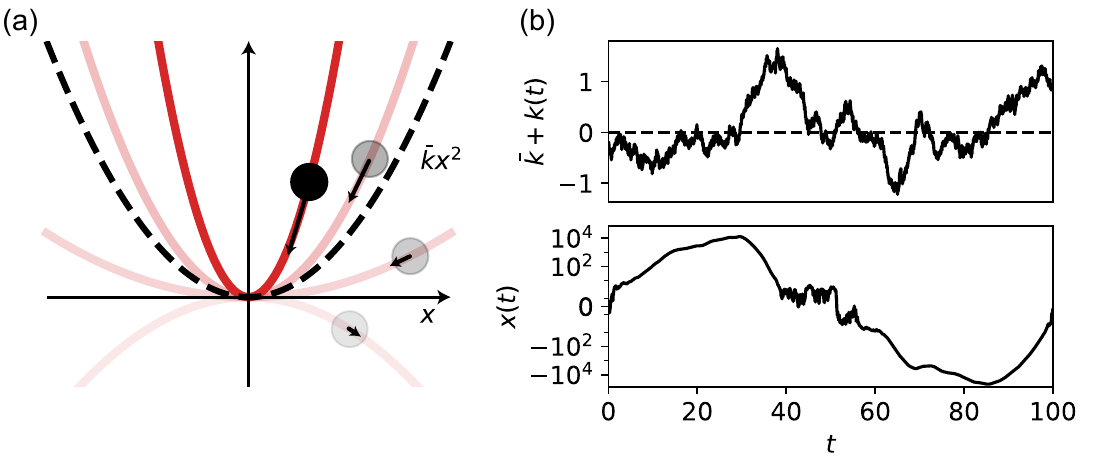}
    \caption{The OU$^2$ process as a minimal model of dissipative confinement. (a) Gaussian fluctuations around a positive mean in the stiffness coefficient of the harmonic potential acting on an overdamped Brownian particle lead to the establishment of futile breathing cycles, with transients of increasing/decreasing stiffness driving the particle closer to/further from the origin. (b) Example trajectories for the particle position $x(t)$ and the total confining potential stiffness $\bar{k} + k(t)$, with $\bar{k} = 1$, $D_x = 1$, $\mu = 0.01$ and $D_k = 0.02$.}
    \label{fig:schematic}
\end{figure}

Inspired by this rather simple idea, we define here a generic model of diffusion in a noisy trap. Namely, we introduce continuous, zero-mean fluctuations in the potential stiffness of the original OU process [Eq.\,(\ref{eq:original_OU})]. More precisely, we model these fluctuations themselves by an OU process: we characterise this second process by an effective stiffness $\mu$ and effective diffusivity $D_k$ (see Fig.~\ref{fig:schematic} for a schematic illustration and example trajectories). Overall, the resulting coupled dynamics of the particle position $x(t)$ and the fluctuations in the confining potential stiffness $k(t)$ read
\begin{subequations}
\begin{align}
    \frac{d x(t)}{dt} &= - \left[\bar{k} + k(t)\right] x(t) + \sqrt{2 D_x} \zeta_x(t)  \label{eq:OU2_x} \\
    \frac{d k(t)}{dt} &= -\mu k(t) + \sqrt{2D_k} \zeta_k(t) \label{eq:OU2_k}
\end{align}
\label{eq:OU2}
\end{subequations}
where we fix the average stiffness $\bar{k} > 0$, $k(t)$ is the zero-mean fluctuating contribution and $\langle \zeta_i(t) \zeta_j(t) \rangle = \delta_{ij}\delta(t-t')$. The particle dynamics reduce to a standard OU process upon setting $D_k = 0$. We call this generic composite stochastic process the \textit{OU$^2$ process} and dedicate the rest of this paper work to its extensive characterisation\footnote{The OU$^2$ model, which we introduce here, is not to be confused with the squared-OU models, a term sometimes used in the context of the modelling of interest rates by the Cox-Ingersoll-Ross model \cite{Cox1985}.}. To the best of our knowledge, this model has been introduced for the first time in two recent works by the authors \cite{Alston2022,Cocconi2023} in the context of nonequilibrium thermodynamics of diffusion in fluctuating potentials. Alternative generalisations of the OU process have also been investigated including models in which the location of the confining potential minimum undergoes stochastic \cite{Manikandan2017} or oscillatory \cite{Lee2023} ``sliding'' dynamics at fixed stiffness. The OU$^2$ model is closely related to a number of other models, which we mention briefly below.

%%%%%%%%%%%%%
%%%%%%%%%%%%%

Firstly, we note that the OU$^2$ process is closely related to the problem of Brownian motion in an intermittent harmonic potential \cite{Santra2021,Zhang2017}, where the potential switches stochastically between two states with finite stiffnesses $k_1$ and $k_2$ in the manner of a telegraph process. The case where $k_1 =0$ and $k_2>0$ has recently received some attention as it represents a realistic implementation of stochastic resetting \cite{Evans2020,Gupta2020a,Gupta2020b,Jerez2021,Alston2022}; interestingly, it was shown in this case that any degree of intermittency leads to the establishment of a nonequilibrium stationary probability density for the trapped particle's position displaying a Gaussian bulk which eventually crosses over into exponential tails. 
%\tbr{For the more general case where the stiffness of the potential switches stochastically between two finite values $k_1$ and $k_2$ following a two-state Markov jump process, it is also notable that the condition for existence of a moment of order $s$ is more restrictive than merely ensuring that the stiffness is positive on average, $\langle k \rangle_t >0$. Indeed, the existence of $\langle |x|^s \rangle$ requires that $k_1 P(k_1) + k_2 P(k_2) - s k_1 k_2 < 0$ with $P$ the stationary probability mass function of the jump process \cite{Guyon2004}. We will see that something very similar occurs in the OU$^2$ case.}\tbc{Moved this comment at the start of Section 4} Let us also note here that both models can be related to the dynamics of the displacement in nonequilibrium two-particle systems with intermittent and continuously fluctuating pair interactions \cite{Alston2022,Cocconi2023}. 

Furthermore, the OU$^2$ process introduced here constitutes a continuous time extension of a class of so-called ``random difference equations" (see for instance \cite{Kesten1973}), i.e.\ recurrence relations involving random parameters. For instance, the model introduced in Ref.\,\cite{Morita2016} is effectively a time-discretised version of the OU process where the confining potential exhibits a fluctuating stiffness, whose fluctuations are uncorrelated in time. More recently, Morita \cite{Morita2018} focused on a version of these processes where the random parameter is allowed to have correlations in time. Nevertheless, in this work, the author consider the much simpler case where $k(t)$ is governed by a Poisson jump process, arguing that studying the case where $k(t)$ is governed by an OU process, which is precisely a discrete-time analog of OU$^2$ process, is particularly challenging. 

In the study of transport in inhomogeneous environments, recent models of \textit{diffusing diffusivity} have been introduced, which allow for stochastic fluctuations in diffusivity of a free Brownian particle. These models display Fickian diffusion (characterised by a linear time dependence of the mean-square displacement) in the presence of non-Gaussian displacement distributions \cite{Chubynsky2014,Banks2016,Sposini2018}.

Moreover, establishing a connection with non-equilibrium thermodynamics, the OU$^2$ process can be seen as a stochastically breathing harmonic potential. Standard breathing potentials, whereby the stiffness is modulated deterministically in time according to a pre-defined protocol, are often studied in the context of heat engines operating in finite time cycles \cite{Blickle2012,Martinez2016,Martinez2017}. Interestingly, generic results have been obtained in the slow driving regime for the full distribution of the stochastic work \cite{Speck2011}. 

An alternative inspiration for the OU$^2$ process may be found in the context of motile active matter, particularly in the canonical active particle model known as the Active Ornstein-Uhlenbeck particle \cite{Martin2021,Bothe2021} (AOUP). Here, out-of-equilibrium self-propulsion is introduced via a forcing term in the Langevin equation of motion, whose statistics are those of a zero-mean OU process. However, one may equivalently interpret this term as a \textit{linear} potential whose amplitude is modulated stochastically in time. The OU$^2$ process is thus a natural, non-motile counterpart to the (much more studied) AOUP, offering a minimal example of dissipative trapping and single-particle irreversibility beyond active motility.

Finally, we will see shortly that the so-called \textit{random acceleration model} can be a seen as a special case of the OU$^2$ process \cite{Burkhardt2000,Majumdar2001,Majumdar2010,Boutcheng2016,Singh2020}.\\

The paper is organised as follows: we begin in Section \ref{sec:nonoise} with a preliminary calculation of the marginal probability density function of the position for an OU$^2$ process with vanishing positional noise, $D_x=0$, highlighting some non-trivial characteristics of the associated statistics. In Section \ref{sec:green}, we move away from this limit and derive the conditional and full Green's functions of the full OU$^2$ process, clarifying the condition for the stability of the dynamics.  Section \ref{sec:moments} deals with the moments of the marginal, stationary probability density function for coordinate $x$ in Eq.~\eqref{eq:OU2_x}. In particular, we obtain the necessary conditions for the existence of the even moments $\langle x^{2n}\rangle$ for arbitrary $n \geq 1$ in the form of an upper bound on $\bar{k}\mu^2/D_k$ which decreases monotonically to zero with increasing $n$, as well as closed form expressions for the second and fourth moments. In Section \ref{sec:fastslow}, we study two limits of fast stiffness dynamics by means of homogenisation \cite{Pavliotis2008}, solving analytically the resulting coarse-grained Fokker-Planck equation for the slow dynamics. 
In Section \ref{sec:maxima}, we draw on known heuristic arguments from extreme value statistics of weakly correlated time series to conjecture the distribution of the maximum of a finite time OU$^2$ process, verifying our proposed classification via numerical simulations.
Owing to the algebraic nature of the tails of the marginal probability density, we observe an unexpected transition from the Gumbel to the Frechet universality class. 
Section \ref{sec:fpt} is dedicated to investigating the impact of stiffness fluctuations on the mean first passage time (mFPT) to a stationary target. 
Finally, we re-derive previous results for the steady-state entropy production rate \cite{Alston2022,Cocconi2023} in a compact way in Section \ref{sec:epr}, thus offering a simple thermodynamic characterisation of the model. Finally, some remarks and potential directions for future research are discussed in the Conclusion.

%%%%%%%%%%%%%%%%%%%%%%%%%%%%%%%%%%
% Section: Exact solution in the limit of vanishing positional noise  %
%%%%%%%%%%%%%%%%%%%%%%%%%%%%%%%%%%
\section{Exact solution in the limit of vanishing positional noise}
\label{sec:nonoise}

As a preliminary analysis, we consider the limiting case of vanishing positional noise, $D_x =0$ in Eq.~\eqref{eq:OU2_x}, for which a series of exact results can be derived. In this case the dynamics of the particle are constrained to the half-line to which the initial condition belongs. Here, by symmetry, we set $x(0) > 0$ so that $x(t)>0$ without loss of generality.

%%%%%%%%%%%%%%%%%%%%%%
% Subsection: Probability density function %
%%%%%%%%%%%%%%%%%%%%%%
\subsection{Probability density function}
Let us denote $x_0 \equiv x(0)$ and define $z\equiv\ln x$ such that we can recast the dynamics as
\begin{equation}\label{eq:z_dyn}
    \dot{z}(t) = -\bar{k} - k(t)
\end{equation}
or, equivalently,
\begin{equation}
    \ddot{z}(t) = \mu k(t) - \sqrt{2D_k} \zeta_k(t)~,
\end{equation}
which reduces to the random acceleration process \cite{Burkhardt2000,Majumdar2001,Majumdar2010,Boutcheng2016,Singh2020} in the limit $\mu \to 0$. 

%%%%%%%%%%%%%%%%%%%%%%%
\begin{figure}
\centering
\includegraphics[width=0.9\textwidth]{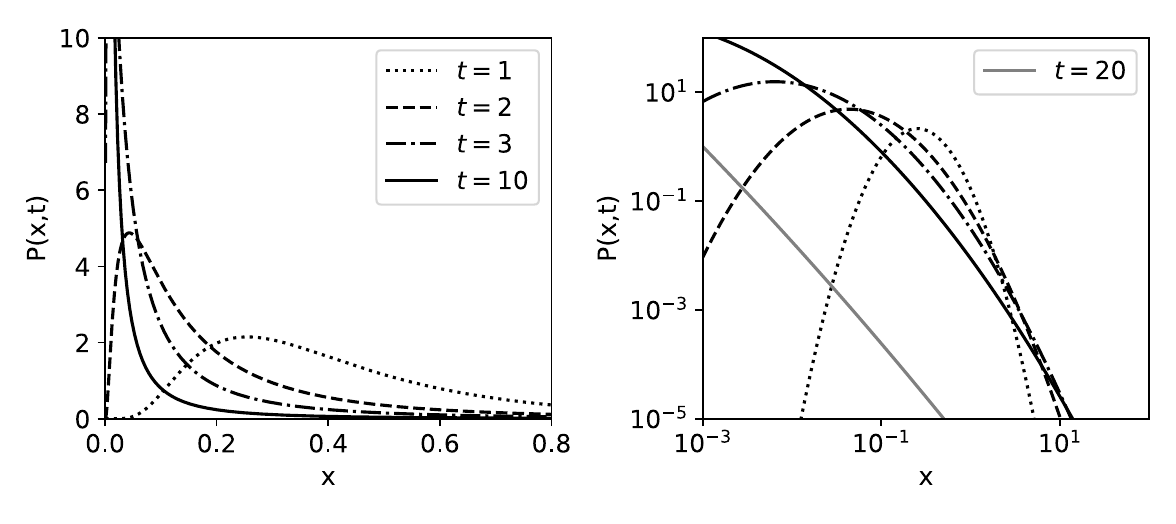}
\caption{Marginal probability density function $P(x,t)$ in the limit of vanishing positional noise, as given by Eq.~\eqref{eq:pdf_x_prel_a}, at various time in linear and double-logarithmic scales. Here, we set $\bar{k}=1$, $x_0=1$ and $2D_k/\mu^2=1$. }
\label{fig:nonoise_pdf}
\end{figure}
%%%%%%%%%%%%%%%%%%%%%%%

As initial condition at $t=0$, we choose $z(0)=z_0$ (and correspondingly, $x_0 = e^{z_0}$) and assume that the fluctuating stiffness has been evolving from $t \to -\infty$ such that at $t=0$ the particle experiences a value of $k$ randomly drawn from the steady-state distribution. The solution of Eq.~\eqref{eq:z_dyn} is thus written
\begin{equation}\label{eq:z_formsol}
    z(t) = z_0 - \bar{k}t - \sqrt{2D_k}\int_0^t dt' \int_{-\infty}^{t'} dt'' e^{-\mu(t'-t'')} \zeta_k(t'')~,
\end{equation}
of which the time-dependent mean and variance are computed straightforwardly to be 
\begin{subequations}
\begin{align}
    \bar{z}(t) &\equiv \langle z(t)\rangle = z_0 - \bar{k}t, \label{eq:exact_average}\\
    \sigma_z^2(t) &\equiv \langle z^2(t) \rangle - \langle z(t)\rangle^2 = \frac{2D_k}{\mu^2} t - \frac{2D_k}{\mu^3}(1-e^{-\mu t})~, \label{eq:exact_variance}
\end{align}
\end{subequations}
where $\langle \cdots \rangle$ means an average over realisations of the noise $\zeta_k$. It is interesting to note that at long times, i.e. $t \gg \mu^{-1}$, the variance scales linearly with time as $\sigma_z^2(t) \simeq 2D_k t/\mu^2$ confirming that the process remains diffusive even in the absence of positional noise. Since \eqref{eq:z_formsol} is Gaussian, the first two moments are sufficient to determine the time-dependent 
 moment generating function,
 \begin{equation}
     Z(q,t) \equiv \langle e^{-iqz(t)}\rangle = \exp\left[ -iq\bar{z}(t) - \frac{\sigma_z^2(t)}{2}q^2 \right]
 \end{equation}
 from which the corresponding probability density is easily obtained,
\begin{equation}\label{eq:sol_zeronoise}
    P(z,t) = \frac{1}{\sqrt{2\pi\sigma^2_z(t)}}\exp\left[-\frac{(z-\bar{z}(t))^2}{2\sigma^2_z(t)}\right].
\end{equation}
Via a straightforward transformation of probability, we then obtain that the probability density for the original variable $x(t)$ is given by the following log-normal distribution 
%\begin{subequations}
\begin{align}
    P(x,t) &= \frac{dz}{dx} P(z(x),t) 
    = \frac{1}{x \sqrt{2\pi\sigma^2_z(t)}}  \exp\left[ - \frac{ (\ln(x) - \bar{z}(t))^2}{2\sigma^2_z(t)} \right], \label{eq:pdf_x_prel_a}
%    & \underrel[c]{t \gg \mu^{-1}}{\simeq} \quad \frac{\mu e^{-\frac{(\mu\bar{k}t)^2}{4D_kt}}}{x_0 \sqrt{4\pi D_k t}}  \exp\left[ - \frac{\mu^2 \ln(x/x_0)^2}{4D_k t} - \left( \frac{\bar{k}\mu^2}{2D_k}+1\right)\ln(x/x_0) \right],
%        & \underrel[c]{t \gg \mu^{-1}}{\simeq} \quad \frac{\mu}{x_0}\frac{ \exp\left[ -\bar{k}^2 \mu^2 t /4D_k\right] }{\sqrt{4 \pi D_k t}} \exp\left[ -\frac{\mu^2 \ln(x/x_0)^2}{4D_k t} - \ln(x/x_0)\left(\frac{\mu^2 \bar{k}}{2D_k} + 1\right) \right] , \label{eq:pdf_x_prel_b} \\
 %       & \underrel[c]{t \gg \mu^{-1}}{\simeq} \quad \frac{\mu}{x_0}\frac{\exp\left[ -\bar{k}^2 \mu^2 t/4D_k\right]}{\sqrt{4 \pi D_k t}} \exp\left[ -\frac{\mu^2 \ln(x/x_0)^2}{4D_k t}\right] \left(\frac{x}{x_0}\right)^{-(1+\mu^2 \bar{k}/2D_k)}. \label{eq:pdf_x_prel_c}
\end{align}
%\end{subequations}
The exact probability density \eqref{eq:pdf_x_prel_a} is shown for different values of $t$ in Fig.~\ref{fig:nonoise_pdf}.

%For fixed large $t$ and moderate $x/x_0$, the second exponential term in \eqref{eq:pdf_x_prel_c} is well approximated by $1$ and the probability density is overall well approximated by the following power law in $x$ 
%\begin{equation}\label{eq:pdf_x_pow}
%    P(x,t) \propto x^{-\left( \frac{\bar{k}\mu^2}{2D_k} + 1 \right)}
%\end{equation}
%which is eventually regularised at $x \to 0$ and $x \to \infty$ by the squared logarithm term, thus ensuring normalisability. Depending on the value of the parameter combination $\bar{k}\mu^2/D_k$, the expectation values will be controlled by one or the other cutoffs. The crossover points $x_{\pm}$ for the appearance of these soft cutoffs can be estimated by determining the value of $x$ at which the magnitude of the linear and quadratic terms in the exponent are equal,
%\begin{equation}
%	\frac{\mu^2 \ln(x_\pm/x_0)}{4D_k t} = \left| \frac{\bar{k}\mu^2}{2D_k}+1 \right|
%\end{equation}
%which gives the exponential time dependence
%\begin{equation}\label{eq:pdf_x_cutoff}
%    x_\pm = x_0 \exp\left( \pm 2 t \left| \bar{k} + \frac{2D_k}{\mu^2} \right| \right)~.
%\end{equation}
%For $x \gg x_+$ and $x \ll x_-$, on the other hand, the probability density decays faster,
%\begin{equation}
%    P(x,t) \propto x^{ - \frac{\mu^2 \ln(x/x_0)}{4D_k t} }~,
%\end{equation}
%albeit still sub-exponentially.

%%%%%%%%%%%%%%%%%%%
% Subsection: Growth and trapping %
%%%%%%%%%%%%%%%%%%%
\subsection{Growth and trapping}

Equipped with the time-dependent probability density \eqref{eq:pdf_x_prel_a}, we can start to study the main qualitative features of the system. Namely, studying three key statistics of the process --- the median, mean and mode  of $x(t)$ --- each of which offers a different perspective on the dynamics, we discuss the conditions under which the process is said to be: (i) \textit{trapped} in the sense that the associated statistic reverts back to the center of the potential (here, $x=0$) or (ii) \textit{growing} in the sense that the associated statistic grows exponentially in time. 

\paragraph{Median ---}
The median is defined as the value $x_M$ in the support of the distribution such that
\begin{equation}
    \int_0^{x_M} dx \ P(x,t) = \frac{1}{2}~.
\end{equation}
For the log-normal distribution \eqref{eq:pdf_x_prel_a}, an explicit expression for the median can be written and simply gives
\begin{equation}
x_M(t) = \exp[\bar{z}(t)] = x_0 \exp(-\bar{k}t)
\label{eq:medappreq}
\end{equation}
Interestingly, we note that the behavior of the median changes as $\bar{k}$ changes sign. Indeed for $\bar{k}>0$, we can easily see that the median $x_M(t)$ approaches zero exponentially as $t$ increases. Conversely, for $\bar{k}<0$, we find that $x_M(t)$ grows exponentially with $t$. Note that the same result holds for every finite percentile of the distribution. In other words, as $\bar{k}$ changes sign from positive to negative, \textit{most of the particles} go from being trapped around $x=0$ to seeing their position grow exponentially in time.  

\paragraph{Mean ---}
Secondly, let us consider the mean of the distribution \eqref{eq:pdf_x_prel_a}, which can be computed exactly to be
\begin{equation}
    \label{eq:mean_x_nonoise}
    \bar{x}(t)= \exp\left[\bar{z}(t) + \frac{\sigma_z^2}{2}\right] = x_0 e^{\frac{D_k}{\mu^3}(1-e^{-\mu t})}\,e^{-\left(\bar{k} - \frac{D_k}{\mu^2}\right)t} \underrel{t\to\infty}{\propto} \exp\left[-\left(\bar{k} -\frac{D_k }{\mu^2}\right)t\right].
\end{equation}
The mean thus decays exponentially with time for $\bar{k} > D_k/\mu^2$, a more stringent condition compared to that of median trapping. 

\paragraph{Mode ---}
Finally, we turn our attention to the mode $x_m$, which is defined as the location of the maximum of the probability density, $\left. \partial_x P(x,t) \right|_{x_m}=0$. For the log-normal distribution in Eq.~\eqref{eq:pdf_x_prel_a}, the mode is given by
\begin{equation}
    \label{eq:mode}
    x_m(t) = \exp\left[\bar{z}(t) - \sigma_z^2 \right] = x_0 e^{-\frac{2D_k}{\mu^3}(1-e^{-\mu t})}\,e^{-\left(\bar{k} +\frac{2D_k }{\mu^2}\right)t} \underrel{t\to\infty}{\propto} \exp\left[-\left(\bar{k} +\frac{2D_k}{\mu^2}\right)t\right].
\end{equation}
Here, we notice that the behavior of the mode for the OU$^2$ process changes from trapped to growing as $\bar{k} + 2D_k/\mu^2$ changes sign.  In other words, for $\bar{k} < -2D_k/\mu^2$, the distance of the \textit{most likely outcome} grows exponentially.

\paragraph{}
Remarkably, these results imply the existence of two non-trivial regimes:
\begin{itemize}
\item[(i)] for $0<\bar{k}<D_k/\mu^2$, the mean position of an ensemble of particles undergoing OU$^2$ processes without positional noise grows indefinitely even if most of the particles approach zero exponentially; this is due to rare trajectories with exceptionally large displacements.
\item[(ii)] for $-2D_k/\mu^2 < \bar{k} < 0$, most of the particles escape exponentially to infinity, however the most likely outcome remains trapped at the proximity of the origin.
\end{itemize}

%%%%%%%%%%%%%%%%%%%%%%%%%
% Subsection: Example of possible intervention %
%%%%%%%%%%%%%%%%%%%%%%%%%
\subsection{Optimal trapping and condition for growth}

It is interesting to note that, while the parameters $\mu$ and $D_k$ might not be accessible to direct experimental control in many physical implementations of this model, one may still be able to control the strength of the couplings between the particle and the potential via a medium-dependent parameters $u$ such that $\dot{x}=-u(\bar{k}+k(t))$. In this case, the mean position becomes $\bar x(t;u) \propto \exp{\left[\left(-\bar{k} u +u^2\frac{D_k}{\mu^2}\right)t\right]}$ for large enough $t$. For example in an experiment with optical tweezers, $u$ can be tuned by changing the viscosity or dielectric properties of the colloid, while it may not be possible to improve the properties of the confining potential by increasing its mean $\bar{k}$ or reducing the noise strength $D_k$.  The exponent is now negative for $u<\frac{\mu^2\bar{k}}{D_k}$. In other words, provided that $\bar{k}$ is positive we can always induce mean trapping by reducing the coupling of the particles with the potential itself. On the other hand, an excessive reduction of the coupling $u$ results in a weak confinement. The optimal value of $u$ minimising the exponent, hence the magnitude of the mean, and thus providing the best confinement of the latter is $u_{\rm opt}={\mu^2}/{2D_k}$. 

Another application of the OU$^2$ process can be found in finance, where Eq.~\eqref{eq:OU2_x} can be used as a simplified model for the growth of the capital $x$ of a company. In this case, exponential growth rather than trapping is the desired outcome. In this model, the stiffness $k$ represents instead (minus) the return on investment of its operations, which is itself subject to stochastic market fluctuations. While the mean return $-\bar{k}$ and volatility $D_k$ of investment can be difficult to improve, $u$ can be tuned by simply reinvesting more capital, while $u>1$ can be obtained by using leverage. The expected capital will grow for $u>u_{\rm{th}}=\frac{\mu^2\bar{k}}{D_k}$, which can happen also for companies with negative mean returns.

%%%%%%%%%%%%%%%%%%%%%%%%%
% Section: Conditional and full Green's function %
%%%%%%%%%%%%%%%%%%%%%%%%%
\section{Conditional and full Green's function}
\label{sec:green}

We now return to the full model, including noise in the displacement, and calculate its Green's function, first conditional on a particular value of $k$ at the time of perturbation, then averaged over the corresponding steady-state distribution. From Eq.\,\eqref{eq:OU2}, we can write that the formal solution for $x$ is given by
\begin{equation} \label{eq:formal_sol}
    x(t) = \sqrt{2D_x} \int_{-\infty}^t dt' \ \zeta_x(t') \exp\left[ - \bar{k}(t-t') - \int_{t'}^t dt'' \ k(t'') \right]~.
\end{equation}
From Eq.\,\eqref{eq:formal_sol}, we identify the \textit{conditional} Green's function of the process, which describes the typical evolution of a noise-generated perturbation,
\begin{equation}
    \mathcal{G}(t;k_0) = \left\langle \exp\left[ -\bar{k}t - \int_0^t dt' \ k(t') \right] \right\rangle_{k_0} \Theta(t)
\end{equation}
where $\langle \bullet \rangle_{k_0}$ denotes an average with respect to the possible realisations of the process $k(t)$ conditioned on the initialisation $k(0) = k_0$, and the Heaviside theta function $\Theta(t)$ ensures causality. The conditional Green's function $\mathcal{G}(t;k_0)$ quantifies the typical temporal evolution of a perturbation generated at $t=0$ by the noise $\zeta_x$. 
Exploiting the relation between the moment and cumulant generating functions and the fact that the cumulants of order 3 and above vanish for the OU process due to it being Gaussian, we write
\begin{subequations}
\begin{align}
    \mathcal{G}(t;k_0) 
    &= e^{-\bar{k}t}  \exp \sum_{n=1}^\infty \frac{(-1)^n}{n!} \left\langle \left( \int_0^t dt' k(t') \right)^n \right\rangle_{c,k_0} \\
    &= e^{-\bar{k}t} \exp \left[ -\int_0^t dt' \langle k(t')\rangle_{c,k_0} + \frac{1}{2} \int_0^t dt' dt'' \ \langle k(t') k(t'') \rangle_{c,k_0} \right]~,
\end{align}
\label{eq:conditional_green}
\end{subequations}
where $\langle \bullet \rangle_{c,k_0}$ denotes the conditional cumulants. The conditional cumulants of first- and second-order can easily be calculated independently 
\begin{subequations}
\begin{align}
    \langle k(t')\rangle_{c,k_0} &= k_0 e^{-\mu t'}, \\
    \langle k(t') k(t'') \rangle_{c,k_0} &= \frac{D_k}{\mu} \left(e^{-\mu|t'-t''|}-e^{-\mu(t'+t'')}\right).
\end{align}
\end{subequations}
With this result in hand, we can perform the integral in \eqref{eq:conditional_green} to obtain the conditional propagator
\begin{equation} \label{eq:cond_prop}
    \mathcal{G}(t;k_0) = \exp\left[ -\left(\bar{k} - \frac{D_k}{\mu^2} \right)t - \frac{k_0}{\mu}(1-e^{-\mu t}) + \frac{D_k}{2\mu^3}(4 e^{-\mu t}- e^{-2\mu t} - 3) \right]~.
\end{equation}
Interestingly, the propagator decays to zero at long times only if $\bar{k} > D_k/\mu^2$, while fluctuations grow exponentially otherwise. This highlights the importance of the competition between the two timescales in the problem, namely $\tau_x = 1/ \bar{k}$ --- the typical mean reversion time for the particle position --- and $\tau_k = \mu^2/D_k$ --- the typical timescale for the stiffness fluctuations.

Also note that the dependence on the initial condition for the stiffness, $k_0$, is rather simple. Expanding the exponent to leading order in small times $t \ll 1$, we find
\begin{equation}
    \mathcal{G}(t;k_0) = \exp[-(\bar{k}+k_0)t + \mathcal{O}(t^2)]
\end{equation}
indicating that at short times the growth/decay of fluctuations is controlled by the initial condition $k_0$, such that fluctuations might initially grow exponentially (when $k_0 < - \bar{k}$), even when they are eventually suppressed on average at long times. In other words, the conditional propagator is not necessarily monotonic. 

We might also be interested in a situations where the initial value of the potential stiffness $k_0$ is unknown. Assuming that the statistics of $k(t)$ have reached steady-state by the time we perturb our system, we can calculate the full propagator by averaging Eq.~\eqref{eq:cond_prop} over $k_0$ which has a known steady-state Gaussian probability density function, i.e.\
\begin{equation}
    \mathcal{G}_{\rm full}(t) = 
    \exp\left[ -\left(\bar{k} - \frac{D_k}{\mu^2} \right)t + \frac{D_k}{2\mu^3}(4 e^{-\mu t}-  e^{-2\mu t} - 3) \right] \left\langle \exp \left[ - \frac{k_0}{\mu}(1-e^{-\mu t}) \right] \right\rangle 
\end{equation}
where $\langle \bullet \rangle$ now denotes an expectation with respect to the steady-state probability density function of $k_0$. Using the fact that the last term in the above is simply the moment generating function of a zero-mean normal distribution with conjugate variable $s = -(1-e^{-\mu t})/\mu$, we eventually arrive at
\begin{align}
    \mathcal{G}_{\rm full}(t) 
    &= \exp\left[ -\left(\bar{k} - \frac{D_k}{\mu^2} \right)t + \frac{D_k}{2\mu^3}(4 e^{-\mu t}- e^{-2\mu t} - 3) + \frac{D_k}{2\mu^3} (1-e^{-\mu t})^2 \right] \nonumber \\
    &= \exp\left[ -\left(\bar{k} - \frac{D_k}{\mu^2} \right)t + \frac{D_k}{\mu^3}(e^{-\mu t}-   1) \right]~. \label{eq:full_green}
\end{align}
We note that in the limit where $D_k \to 0$, we recover the Green's function for the standard OU process, c.f.~Eq.~\eqref{eq:OU_prop}. The long time behaviour is the same as for the conditional case, however expanding again at small times $t \ll 1$, we now find $\mathcal{G}_{\rm full}(t)  = \exp\left[-\bar{k}t + \mathcal{O}(t^2) \right]$. This is to be expected since the average of the exponential converges to the exponential of the average when $t \to 0$. 

%%%%%%%%%%%%%%%%%
% Section: Positional moments  %
%%%%%%%%%%%%%%%%%
\section{Positional moments}
\label{sec:moments}

We now discuss the steady-state moments of the marginal probability density function for the coordinate $x$. In particular, we derive conditions for the existence of finite moments. Already in the case of a Brownian particle confined in a potential whose stiffness switches stochastically between two finite values $k_1$ and $k_2$ following a two-state Markov jump process, it was shown that the condition for existence of a moment of order $s$ is more restrictive than merely ensuring that the stiffness is positive on average, $\langle k \rangle_t >0$. Indeed, the existence of $\langle |x|^s \rangle$ requires that $k_1 P(k_1) + k_2 P(k_2) - s k_1 k_2 < 0$ with $P$ the stationary probability mass function of the jump process \cite{Guyon2004}. We will see in this section that similar conditions can be derived in the OU$^2$ case.

Starting from the formal solution, Eq.~\eqref{eq:formal_sol}, we first argue by symmetry that all odd moments are expected to vanish, $\langle x^{2n+1}(t) \rangle=0$ for all $n \in \mathbb{N}$. The even moments are on the other hand given by
\begin{align} \label{eq:def_evenmom}
    \langle x^{2n}(t) \rangle = (2D_x)^{n} 
    \left\langle \left( \int_{-\infty}^t dt' \ \zeta_x(t') e^{- \bar{k}(t-t')} \exp\left[ - \int_{t'}^t dt'' \ k(t'') \right] \right)^{2n} \right\rangle~.
\end{align}
Using the fact that $\zeta_x(t)$ and $k(t)$ are uncorrelated stochastic processes, we write
\begin{align}
\Bigg\langle \prod_{i=1}^{2n} \zeta_x(t_i')\exp \Bigg[-\bar{k}&(t-t_i') - \int_{t_i'}^t dt_i'' k(t_i'')\Bigg] \Bigg\rangle \nonumber \\
    &=
    \left\langle \prod_{j=1}^{2n} \zeta_x(t_j') \right\rangle
    \left\langle \prod_{i=1}^{2n} \exp\left[-\bar{k}(t-t_i') - \int_{t_i'}^t dt_i'' k(t_i'')\right]\right\rangle~.
\end{align}
The first expectation on the right-hand side can be simplified by Wick-Isserlis theorem to 
a sum of product of white noise correlators, i.e.\ Dirac delta functions, with $(2n-1)!! = (2n)!/(2^n n!)$ summands corresponding to all possible pairings $\mathcal{P}_{2n}$ of the random variables:
\begin{equation}
\left\langle \prod_{j=1}^{2n} \zeta_x(t_j') \right\rangle = \sum_{p \in \mathcal{P}_{2n}} \prod_{\{i,j\} \in p} \left\langle \zeta_x(t'_i) \zeta_x(t'_j) \right\rangle~.
\end{equation}
Since the overall integral is invariant under permutation of the indices, all summands give the same contribution. The expression for the moments thus simplifies to
\begin{equation}
    \langle x^{2n}(t) \rangle = \mathcal{N}_n\int_{-\infty}^t dt_{1<...<n} \  \exp\left[{-2\bar{k}\sum_{i=1}^n (t-t_i)}\right] \left\langle \exp\left[- 2 \sum_{j=1}^n \int_{t_i}^t dt_i' k(t_i')\right]\right\rangle \label{eq:gen_mom_contracted}
\end{equation}
with $\mathcal{N}_n = (2D_x)^{n} (2n-1)!! n!$,
where we have additionally imposed the arbitrary time ordering $t_1 < t_2 < ... < t_n$ in the multiple integrals, compensated by the combinatorial prefactor $n!$, without loss of generality. 
Next, we exploit the identity relating the moment generating function and the exponential of the corresponding cumulant generating function
\begin{align}
    \Bigg\langle \exp\Bigg[ - 2 \sum_{i=1}^{n}& \int_{t_i}^t dt_i' \ k(t_i') \Bigg] \Bigg\rangle
    \nonumber \\
    &= \exp \sum_{m=1}^\infty \frac{1}{m!} \left\langle \left( - 2 \sum_{i=1}^{n} \int_{t_i}^t dt_i' \ k(t_i') \right)^m \right\rangle_c \nonumber \\
    &= \exp \left\langle 2 \left( \int_{t_1}^t dt_1' \ k(t_1') + \int_{t_2}^t dt_2' \ k(t_2') + ... + \int_{t_n}^t dt_n' \ k(t_n') \right)^2 \right\rangle_c \label{eq:cum_gen_id}
\end{align}
where we have used the fact that cumulants of order $m>2$ vanish for the equilibrium OU process governing $k(t)$, Eq.~\eqref{eq:OU2_k}, while the first order cumulant is zero at steady state. 
The right-hand side of \eqref{eq:cum_gen_id} can be evaluated as
\begin{align}
\Bigg\langle 2 \Big( \int_{t_1}^t & dt_1' \ k(t_1') + \int_{t_2}^t dt_2' \ k(t_2') + ... + \int_{t_n}^t dt_n' \ k(t_n') \Bigg)^2 \Bigg\rangle_c \nonumber \\
    &= \frac{2 D_k}{\mu} \left[ 
    \left(\sum_{i=1}^n
    \int_{t_1}^t dt_i' \int_{t_1}^t dt_i'' \ e^{-\mu|t_i'-t_i''|} \right)
    +
    2 \left( \sum_{i < j}^n \int_{t_i}^t dt_i' \int_{t_j}^t dt_j' \ e^{-\mu|t_i'-t_j'|} 
    \right)\right] \nonumber \\
    &= \frac{4 D_k}{\mu^3} \left[ 
    \left(\sum_{i=1}^n
    \mu(t-t_i) -1 + e^{-\mu(t-t_i)} \right) \right. \nonumber  \\
    &\quad \qquad \qquad \left. +
    2 \left( \sum_{i < j}^n e^{-\mu(t-t_i)} + e^{-\mu(t-t_j)} - e^{-\mu|t_j-t_i|} + 2\mu(t-t_j) - 1 
    \right)\right]~. \label{eq:expansion_cum_gen}
\end{align}
Here we have used the result for the double integral
\begin{align}
    \int_{t_i}^t dt_i' \int_{t_j}^t dt_j' \ e^{-\mu|t_i'-t_j'|} 
    &= \frac{1}{\mu^2} \left( e^{-\mu(t-t_i)} + e^{-\mu(t-t_j)} - e^{-\mu|t_j-t_i|} + 2\mu(t-{\rm max}(t_i,t_j)) - 1\right)~.
\end{align}
%which reduces to
%\begin{equation}
%    \int_{t_i}^t dt_i' \int_{t_i}^t dt_i'' \ e^{-\mu|t_i'-t_i''|} = \frac{2[\mu(t-t_i) -1 + e^{-\mu(t-t_i)}] }{\mu^2}  
%\end{equation}
%for the limiting case $t_i = t_j$, together with the fact that, by time ordering and within the double sum with $j>i$, it is always the case that $t_j > t_i$ and thus ${\rm max}(t_i,t_j) = t_j$.
We now compute \eqref{eq:expansion_cum_gen} for different values of $n$.

%%%%%%%%%%%%%%%%%%
% Subsection: Variance ($n=1$)   %
%%%%%%%%%%%%%%%%%%
\subsection{Variance ($n=1$)}
\begin{figure}
    \centering
    \includegraphics[scale=0.75]{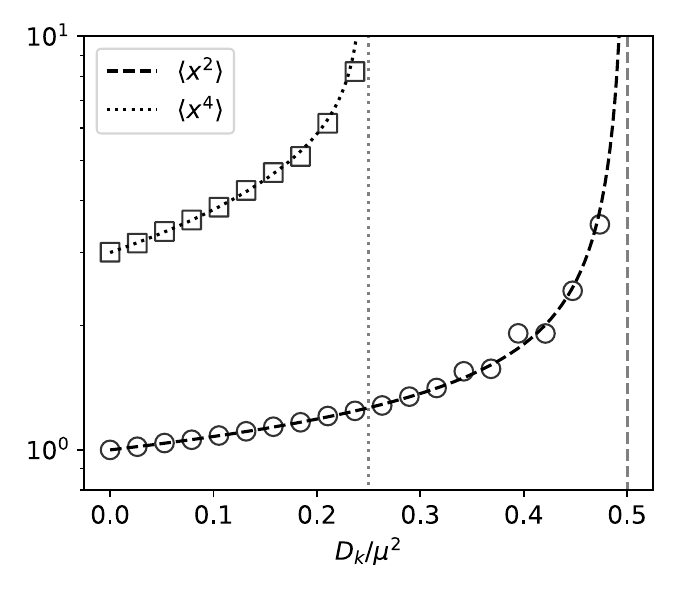}
    \caption{Comparison between analytical results and numerical simulations for the second and quartic steady-state moments of the marginal probability density function for the coordinate $x$ of the OU$^2$ process, showing good agreement between the two. The analytical result for the second moment $\langle x^2 \rangle$, given by Eq.~\eqref{eq:mom2_sum_form} is defined for $D_k/\mu^2 < \bar{k}/2$, while that for the quartic moment $\langle x^4 \rangle$, given by Eq.~\eqref{eq:mom4_sumform} is defined for $D_k/\mu^2 < \bar{k}/4$. Here, we set $D_x =1$ and $\bar{k}=1$. Grey vertical lines indicate the predicted radius of convergence.}
    \label{fig:moments_check}
\end{figure}

First consider the particular case $n=1$ for which compact expressions can be obtained. In this case the right-hand side of \eqref{eq:cum_gen_id} using \eqref{eq:expansion_cum_gen} simplifies to
\begin{equation}
    \left. \left\langle \exp\left[ - 2 \sum_{i=1}^{n} \int_{t_i}^t dt_i' \ k(t_i') \right] \right\rangle \right|_{n=1} = \exp\left[ \frac{4D_k}{\mu^3} \left( e^{-\mu(t-t_1)} + \mu(t-t_1) -1\right) \right]~.
\end{equation}
Using this result, we can then rewrite \eqref{eq:gen_mom_contracted} for $n=1$ as
\begin{equation} \label{eq:m2_simplified}
    \langle x^2 \rangle = 2D_x \int_{-\infty}^t dt_1 \exp\left[ -  \left( 2\bar{k} - \frac{4D_k}{\mu^2}\right)(t-t_1) + \frac{4D_k}{\mu^3}\left( e^{-\mu(t-t_1)} - 1 \right) \right]~.
\end{equation}
It is clear by inspection that the second moment exists if and only if ${D_k}/{\mu^2} < {\bar{k}}/{2}$.
Note that this is a stricter condition compared to that found in Sec.~\ref{sec:green} for the exponential decay of the Green's function, suggesting the existence of parameter regions for which the steady state exists but not the variance. 
We now write the double exponential term in Eq.~\eqref{eq:m2_simplified} as a power series,
\begin{equation}
    \exp\left[ \frac{4D_k}{\mu^3} e^{-\mu(t-t_1)}\right]  = \sum_{\ell=0}^\infty \frac{1}{\ell!}  \left( \frac{4D_k}{\mu^3} \right)^\ell e^{-\mu \ell(t-t_1)}~.
\end{equation}
Substituting back into \eqref{eq:m2_simplified}, swapping integral and sum, performing the simple exponential integral and rearranging terms, we eventually arrive at the expression 
\begin{equation}
    \langle x^2 \rangle  = \frac{2D_x}{\mu} e^{-\xi} \sum_{\ell=0}^\infty \frac{\xi^\ell}{\ell! (\sigma - \xi+\ell)} \label{eq:mom2_sum_form}
\end{equation}
with $\xi = {4D_k}/{\mu^3}$ and $\sigma = {2\bar{k}}/{\mu} $, which reduces to $\langle x^2 \rangle = D_x/\bar{k}$ for $D_k=0$, as expected of the standard OU process. This analytical result is plotted against numerical simulation in Fig.~\ref{fig:moments_check}, showing good agreement. Formally, the right hand side of Eq.~\eqref{eq:mom2_sum_form} can be written more compactly in terms of the lower incomplete Gamma function $\gamma(a,b)$, which has the following series expansion \cite{Arfken2011}
\begin{equation}
    \gamma(a,b) =  b^{a} \sum_{\ell=0}^\infty \frac{(-b)^\ell}{\ell! (a+\ell)} ~, 
\end{equation}
allowing us to reduce Eq.~\eqref{eq:mom2_sum_form} to $\langle x^2 \rangle = 2D_x  e^{-\xi}  (-\xi)^{-\sigma+\xi}  \gamma(\sigma-\xi,-\xi)$ or, in the original notation,
\begin{equation}
\langle x^2 \rangle = 2 D_x \left(-\frac{4D_k}{\mu^3}\right)^{\frac{4}{\mu}\left(\frac{D_k}{\mu^2} - \frac{\bar{k}}{2}\right)} \exp\left[-\frac{4D_k}{\mu^3} \right]  \gamma\left(\frac{4}{\mu}\left(\frac{\bar{k}}{2} - \frac{D_k}{\mu^2} \right),-\frac{4D_k}{\mu^3}\right)
\end{equation}

%%%%%%%%%%%%%%%%%%%%%
% Subsection: Quartic moment ($n=2$) %
%%%%%%%%%%%%%%%%%%%%%
\subsection{Quartic moment ($n=2$)}

In this case the right-hand side of \eqref{eq:cum_gen_id} becomes, for $n=2$ and using \eqref{eq:expansion_cum_gen},
\begin{align}
    \langle x^4 \rangle = 6 (2D_x)^2 &\int_{-\infty}^t dt_1 \int_{t_1}^t dt_2 \  e^{-2[\bar{k}(t-t_1)+\bar{k}(t-t_2)]} \nonumber \\
    \times \exp\Bigg[ &\frac{4D_k}{\mu^3} \left( 2e^{-\mu(t-t_1)} + 2e^{-\mu(t-t_2)} - e^{-\mu(t_2-t_1)} - 3 + \mu(t-t_1) + 3\mu(t-t_2) \right) \Bigg]~.
\end{align}
Expanding once again the double exponential as power series we then obtain
\begin{align}
    \langle x^4 \rangle =& 6 (2D_x)^2 e^{-\frac{12D_k}{\mu^3}} \sum_{n,m,\ell=0}^\infty \frac{\left( -\frac{1}{2} \right)^k  \left( \frac{8D_k}{\mu^3}\right)^{m+\ell+k}}{m! k! \ell!} \nonumber \\
    &\times\int_{-\infty}^t dt_1 \int_{t_1}^t dt_2 \ \exp\Bigg[ -2\left[\bar{k} - \frac{2D_k}{\mu^2}(t-t_1)  - 2(\bar{k}-\frac{6D_k}{\mu^2})(t-t_2)\right] \nonumber \\
    & \qquad \qquad \qquad \quad \qquad - \mu m (t-t_1) - \mu \ell(t-t_2) - \mu k (t_2 - t_1) \Bigg]~,
\end{align}
indicating that the quartic moment exists if and only if ${D_k}/{\mu^2} < {\bar{k}}/{4}$.
The double integral can now be performed in closed form,
\begin{equation} \label{eq:mom4_sumform}
    \langle x^4 \rangle = 24 D_x^2 e^{-\frac{12D_k}{\mu^3}} \sum_{k,m,\ell=0}^\infty \frac{\left( -\frac{1}{2} \right)^k  \left( \frac{8D_k}{\mu^3}\right)^{m+\ell+k}}{m! k! \ell! \left( \frac{4D_k}{\mu^2}-2\bar{k}-\mu(m+k) \right)\left( \frac{16D_k}{\mu^2}-4\bar{k}-\mu(m+\ell) \right)} ~,
\end{equation}
giving us the most compact expression for the quartic moment. This is plotted against numerical simulations in Fig.\,\ref{fig:moments_check}, showing good agreement in the domain of convergence.

%%%%%%%%%%%%%%%%%%%
% Subsection: Higher values of $n$ %
%%%%%%%%%%%%%%%%%%%
\subsection{Higher values of $n$}

For the general case $n>2$, 
%we don't expect to find closed form expressions and thus 
we focus on determining the criteria for convergence of the moments. As shown in \ref{sec:convergence}, one can generalise the arguments developed above and obtain the following criteria of convergence for the moment of order $2n$
\begin{equation}\label{eq:mom_cov_cont_main}
\frac{D_k}{\mu^2}<\frac{\bar k}{2n}
\end{equation}
which is in agreement with the results we just obtained for the particular cases $n=1$ and $n=2$.

%%%%%%%%%%%%%%%%%%%
\begin{figure}
    \centering
    \includegraphics[scale=0.75]{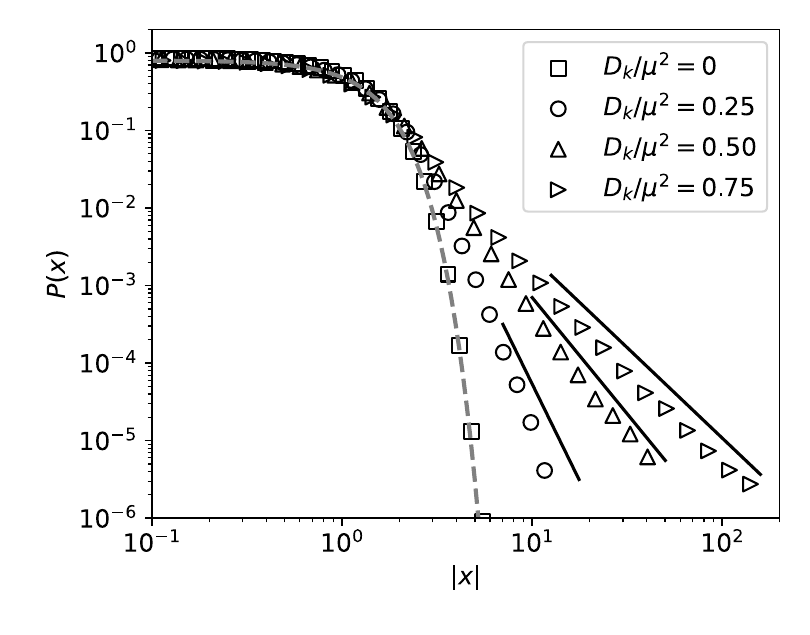}
    \caption{Marginal probability density functions for the coordinate $x$ of the OU$^2$ process obtained from numerical simulations. The asymptotic scaling of the empirical probability is studied as a function of the dimensionless parameter combination $D_k/\mu^2$ controlling the convergence of the marginal moments, as studied in Sec.~\ref{sec:moments}, showing good agreement with the exponents stated in Eq.~\eqref{eq:asymp_scaling_conj}, solid lines. For $D_k/\mu^2=0$, corresponding to the standard OU limit, the distribution is a simple Gaussian (dashed gray line). For any finite $D_k/\mu^2 > 0$, we instead observe a transition to algebraic scaling.
    Here, we work in units such that $D_x = \bar{k} = 1$ without loss of generality.  }
    \label{fig:tails_power}
\end{figure}
%%%%%%%%%%%%%%%%%%%

From this criterion of convergence, we can argue for the asymptotic scaling of the marginal probability density function $P(x)$. Indeed, assuming that the asymptotic scaling exponent is a continuous function of $D_k/\mu^2$, we expect that the value of $D_k/\mu^2$ at which the moment of order $2n$ becomes divergent, namely $D_k/\mu^2 = \bar{k}/(2n)$, corresponds to that at which the marginal probability density scales asymptotically as $x^{-2n-1}$ at large $x$. We conclude that, asymptotically,
\begin{equation}\label{eq:asymp_scaling_conj}
    P(x) \sim |x|^{-1-\frac{\bar{k} \mu^2}{D_k}}~.
\end{equation}
This result in agreement with the closed form expression for $P(x)$ derived analytically in Sec.~\ref{sec:fastslow} below by means of homogenisation in the fast-slow limit, as well as with numerical simulations for the full model, as shown in Fig.~\ref{fig:tails_power}. 

%%%%%%%%%%%%%%%%
% Section: Fast stiffness limit %
%%%%%%%%%%%%%%%%
\section{Fast stiffness limit}
\label{sec:fastslow}

In many applications, such as the optical tweezer example discussed in the introduction, the potential stiffness fluctuations can be reasonably assumed to occur on a comparatively fast timescale.
In this regime, analytical results for the marginal probability density function $P(x)$ can be obtained by enforcing a formal separation of timescales between the \textit{slow} dynamics of the particle position $x(t)$ and the \textit{fast} dynamics of the stiffness fluctuations $k(t)$. The elimination of the fast stiffness dynamics can subsequently be carried out following a multiscale approach \cite{Pavliotis2008}. %, as reviewed in \ref{sec:homogen_review}.
In the following, we consider two fast-slow regimes: (i) a na{\"i}ve adiabatic limit, where the characteristic timescale of the $k$ dynamics is sent to zero at fixed variance $\sigma_k^2 = D_k/\mu$, and (ii) a nontrivial limit, where the variance $\sigma_k^2$ diverges as the inverse of the characteristic timescale, leading to $k(t)$ in Eq.~\eqref{eq:OU2_x} becoming statistically equivalent to a Gaussian white noise. 

%%%%%%%%%%%%%%%%%%%
% Subsection: Naive adiabatic limit  %
%%%%%%%%%%%%%%%%%%%
\subsection{Na{\"i}ve adiabatic limit}
\label{sec:adiabatic}

First, we consider a na{\"i}ve adiabatic limit, in which stiffness fluctuations are expected to be irrelevant. Formally, we introduce a real dimensionless coefficient $\varepsilon$ and proceed to the rescaling $\mu \to \tilde{\mu}/\varepsilon^2$ and $D_k \to \tilde{D}_k/\varepsilon^2$ in Eq.~\eqref{eq:OU2_k} which governs the dynamics of $k(t)$. We then take the limit $\varepsilon \to 0$, keeping the variance of the stiffness $\sigma_k^2 = D_k/\mu = \tilde{D}_k/\tilde{\mu}$ constant. In this limit, one finds that the effective dynamics of the slow variable $x_\varepsilon$ are obtained by replacing $k(t) \to 0$ in Eq.~\eqref{eq:OU2_k} by its mean value. Consequently, $x_\varepsilon$ is shown to be governed by the OU dynamics,
\begin{equation}
	\dot{x}_\varepsilon(t) = - \bar{k} x_\varepsilon(t) + \sqrt{2D_x}\zeta_x(t),
\end{equation}
A rigorous derivation of this result is presented in \ref{sec:homogen_OU2}. We conclude that in this trivial limit, we do not retain any signature of the stiffness fluctuations.

%%%%%%%%%%%%%%%%%
% Subsection: White noise limit %
%%%%%%%%%%%%%%%%%
\subsection{White noise limit}
\label{sec:whitenoise}

Intuitively, we can go beyond this first trivial limit by replacing the Gaussian process $k(t)$ not by its mean value but by a Gaussian white noise with appropriate mean and standard deviation. Formally, this second regime is obtained by performing the alternative rescaling $\mu \to \tilde{\mu}/\varepsilon^2$ and $D_k \to \tilde{D}_k/\varepsilon^4$, before taking the limit $\varepsilon\to 0$ which keeps the ratio $D_k/\mu^2$ constant.
In situations where $x(t)$ denotes the position of an overdamped particle, this regime can be understood physically as a low viscosity, high temperature limit. Indeed, given an effective friction coefficient $\gamma$ and bath temperature $T$, we have by the Stokes-Einstein relation that $\mu \propto \gamma^{-1}$ while $D_x \propto \gamma^{-1}T$. Taking $\gamma = \tilde{\gamma}\varepsilon^{2}$ and $T = \tilde{T} \varepsilon^{-2}$ produces the desired rescaling. 

Mathematically, this amounts to $k(t)$ in Eq.~\eqref{eq:OU2_k} becoming statistically equivalent to a Gaussian white noise with covariance $\langle k(t)k(t')\rangle = 2\tilde{D}_k/\tilde{\mu}^2 \delta(t-t')$. Importantly, the term $k(t)x(t)$ appearing when integrating Eq.~\eqref{eq:OU2_x} should now be treated as a Stratonovich product \cite{Pavliotis2008,dePirey2023}. Taking care of the Stratonovich-to-It{\^o} conversion \cite{Gardiner1985}, we find that the It{\^o} form of the resulting Langevin equation in the limit $\varepsilon \to 0$ reads
\begin{equation} \label{eq:fast_slow_lim_x}
\dot{x}_\varepsilon(t) = - \left(\bar{k} - \frac{D_k}{\mu^2} \right)x_\varepsilon(t)  + \sqrt{2\left(D_x + \frac{D_k x_\varepsilon^2(t)}{\mu^2} \right)} \zeta_x(t)
\end{equation}
where we have used that ${\tilde{D}_k}/{\tilde{\mu}^2} = {D_k}/{\mu^2}$ in this case. Interestingly, this shows a clear instability as $D_k/\mu^2 > \bar{k}$ due to a renormalisation of the confining potential stiffness and, unlike the original dynamics, is characterised by multiplicative noise. The Fokker-Planck representation of Eq.~\eqref{eq:fast_slow_lim_x},
\begin{equation}\label{eq:fp_fastslow_x}
    \partial_t P(x_\varepsilon,t) = \partial_{x_\varepsilon} \left\{ \partial_{x_\varepsilon} \left[ \left( D_x + \frac{D_k x_\varepsilon^2}{\mu^2} \right)P(x_\varepsilon,t)\right] +  \left(\bar{k} - \frac{D_k}{\mu^2} \right) x_\varepsilon P(x_\varepsilon,t) \right\}~,
\end{equation}
can equivalently be derived by multiscale methods (see \ref{sec:homogen_OU2}). Interestingly, Eq.~\eqref{eq:fp_fastslow_x} can be mapped onto an associated Legendre differential equation, see \ref{sec:ass_legendre}. In the rest of this subsection we drop the subscript of $\varepsilon$ and define the shorthands $h \equiv D_k/\mu^2$ and $\kappa \equiv \bar{k}-h$ for the sake of simplicity.

We now proceed to determining the steady state probability density function $P(x)$ associated with the Langevin equation \eqref{eq:fast_slow_lim_x} for the slow $x$ dynamics. To do so, we introduce a variable $z(x)$ whose stochastic dynamics do not involve multiplicative noise \cite{Rubin2014}
\begin{equation}
    z(x) =  \int^x d\xi \left( D_x + \frac{D_k {\xi}^2}{\mu^2} \right)^{-\frac{1}{2}} = \frac{1}{\sqrt{h}}\tanh^{-1}\left(\frac{\sqrt{h}x}{\sqrt{D_x+hx^2}}\right).
\end{equation}
By It{\^o}'s lemma, the dynamics for $z$ take the form
\begin{align}
    \dot{z}(t) = -\frac{(\kappa+h)x}{\sqrt{(D_x+h x^2)}}+ \sqrt{2}\zeta_x(t) = -\frac{(\kappa+h)\tanh\left[\sqrt{h}z(t)\right]}{\sqrt{h}} + \sqrt{2}\zeta_x(t)
\end{align}
where the noise is now additive as we anticipated.

The dynamics for $z$ are exactly those of a passive Brownian particle in a static potential
\begin{equation}
    V(z) = \left( \frac{\kappa + h}{h} \right)\ln \left[\cosh\left(\sqrt{h}z\right)\right]~,
\end{equation}
whence the steady state probability distribution for $z$ is given by the Boltzmann measure
\begin{equation}
    P_Z(z) = e^{-V(z)}/\mathcal{Z},\quad \mathcal{Z} = \sqrt{\frac{\pi}{h}}\frac{\Gamma[(h+\kappa)/2h]}{\Gamma[(2h+\kappa)/2h]}~.
\end{equation}
Finally, we perform a transformation of probability distributions to obtain a simple expression for the probability density of $x$,
\begin{equation} \label{eq:analyt_fastslow_sol}
    P_X(x) = \frac{dz}{dx} P_Z(z(x)) = \sqrt{\frac{h}{\pi}}  \Gamma \left(\frac{\kappa}{2 h}+1\right) \left[\Gamma \left(\frac{h+\kappa}{2 h}\right)\right]^{-1} D_x^{\frac{h+\kappa}{2 h}} \left(D_x+h x^2\right)^{-\frac{\kappa}{2 h}-1} ~.
\end{equation}
Asymptotically, we again find that 
\begin{equation}
P_X(x) \sim |x|^{-\bar{k}\mu^2/D_k-1}
\end{equation}
It can be checked by direct substitution that $P(x)=P_X(x)$ solves the Fokker-Planck equation \eqref{eq:fp_fastslow_x} at steady state.
Notice also that Eq.~\eqref{eq:analyt_fastslow_sol} is in agreement with the asymptotic scaling of the marginal probability density for the full OU$^2$ process, Eq.~\eqref{eq:asymp_scaling_conj}. In particular, it is straightforward to check that the existence of moments of order $2n$ demands $D_k/\mu^2 < \bar{k}/(2n)$. 

\section{Statistics of maxima}
\label{sec:maxima}

Despite their relevance in many domains, such as climate modelling, there is currently no general framework to study the extreme value statistics (EVS) of correlated random variables \cite{Majumdar2020}. Amongst the few exceptions is the standard OU process, for which the EVS can be compute exactly and can be shown to belong to the Gumbel universality class \cite{Haan2006}. Accordingly, the asymptotic probability density of the maximum $X(t) \equiv\, \underrel{\tau \in [0,t]}{\max}\{x(\tau)\}$ is given by
\begin{equation}\label{eq:gumbel}
    \Phi_G(X;m,s) = s^{-1} e^{-(z+e^{-z})}~, \qquad  z = \frac{X-m}{s}
\end{equation}
where the first two moments of $X(t)$ can be expressed as
\begin{equation}
    \langle X\rangle = m + s \gamma, \quad \langle X^2\rangle = \frac{\pi^2 s^2}{6}
\end{equation}
respectively, where $\gamma=0.5772...$ is the Euler-Mascheroni constant. It is thus natural to wonder how the EVS of the OU$^2$ process compare to those of the standard OU process. 

\begin{figure}
    \centering
    \includegraphics[scale=0.7]{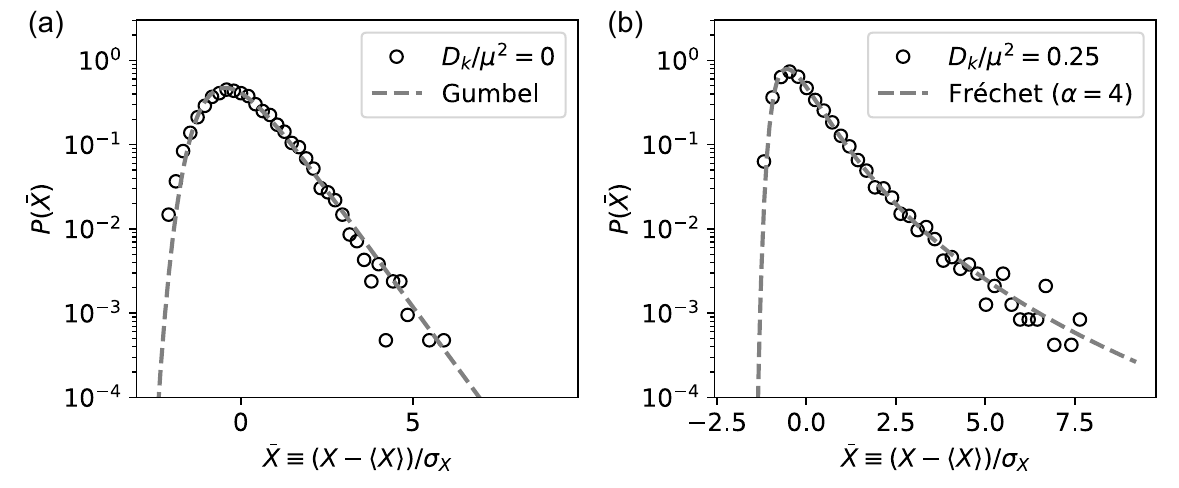}
    \caption{Extreme value statistics for the OU$^2$ process. In the absence of potential fluctuations, $D_k=0$, we recover the standard OU process, for which the limiting distribution of the maximum is known to be of the Gumbel type, Eq.~\eqref{eq:gumbel} (a). In the presence of potential fluctuations, $D_k >0$, we conjecture a transition from Gumbel to Fr\'{e}chet universality class, Eq.~\eqref{eq:frechet}, with a non-universal exponent $\alpha \equiv \bar{k}\mu^2/D_k$ (b). In both cases, we find good agreement between numerical and analytical standardised distributions.}
    \label{fig:f3}
\end{figure}

While a fully analytical approach is beyond the scope of this work, we can draw on the renormalisation group heuristic introduced in Ref.\,\cite{Majumdar2020} to argue that, as long as correlations decay over a finite time, the EVS for a weakly correlated stochastic process are still expected to converge to one of the three limiting distributions for uncorrelated random variables, namely Gumbel, for exponentially decaying parent distributions, Fr\'{e}chet, for fat-tailed parent distributions, and Weibull, for parent distributions with compact support \cite{Majumdar2020,Haan2006}. Combining this heuristic with the finding of Sec.~\ref{sec:moments} that the probability density of $x$ decays algebraically at large $x$, specifically as $-\ln P(x) \sim 1+\bar{k}\mu^2/D_k \equiv 1 + \alpha$, leads us to conjecture that the EVS for the OU$^2$ process should converge to a Fr\'{e}chet distribution with $D_k$-dependent characteristic exponent. In particular, we consider the Fr\'{e}chet probability density for $X(t)=\,\underrel{\tau \in [0,t]}{\max}\{x(\tau)\}$ given by
\begin{equation}\label{eq:frechet}
    \Phi_F(X;m,s) = 
    \begin{cases}
    0 & \text{for} \quad z\leq m \\
    \frac{\alpha}{s}z^{-1-\alpha} e^{-z^{-\alpha}} & \text{for} \quad z > m
    \end{cases}, \qquad \mbox{with} \quad z = \frac{X-m}{s}
\end{equation}
where the first two moments of $X(t)$ are expressed as
\begin{equation}
    \langle X\rangle = m + s \Gamma\left( 1 - \frac{1}{\alpha} \right), \quad \langle X^2\rangle = s^2 \left[ \Gamma\left( 1 - \frac{2}{\alpha} \right) - \Gamma^2\left( 1 - \frac{1}{\alpha} \right)^2 \right]
\end{equation}
the latter being defined only for $\alpha > 2$. We find this conjecture to be in good agreement with numerical simulations of the full model, as shown in Fig.~\ref{fig:f3}. As expected, Gumbel EVS are recovered upon setting $D_k=0$, i.e. in the absence of potential fluctuations.

%%%%%%%%%%%%%%%%%
%%%%%%%%%%%%%%%%%

%%%%%%%%%%%%%%%%%%%%%
% Section: First-passage time statistics  %
%%%%%%%%%%%%%%%%%%%%%
\section{First-passage time statistics}
\label{sec:fpt}

We now focus on the impact of the continuous fluctuations of the potential stiffness present on the first-passage time statistics of the standard Ornstein-Uhlenbeck process, as characterised in \cite{Santra2021,Abkenar2017}. The problem of finding the mean first-passage times to an absorbing boundary can be mapped onto that of the escape of particle over a fluctuating energy barrier. The rate at which a Brownian particle escapes over an energy barrier is a central problem in statistical physics dating back to Kramers \cite{Kramers1940}, finding applications across disciplines through reaction-rate theory \cite{Hanggi1990}. 

Calculating the escape rate over a fluctuating energy barrier for a Brownian particle has attracted some attention in the past \cite{Reimann1995, Reimann1995b,Stein1990,Doering1992}. At low temperatures, zero-mean fluctuations in the energy barrier height lead to so-called {\it resonant activation} and to a reduction of the mean first-passage time, effectively aiding the escape process \cite{Reimann1995, Reimann1995b,Stein1990}; resonant activation has been observed for general confining potentials \cite{Hanggi1990,Doering1992}. In general, solving for the first-passage time distribution and its moments for the coupled dynamics of the particle position and potential stiffness constitutes a formidable task. To the best of our knowledge, only approximate results can be derived in the limit where the timescale associated with the potential stiffness fluctuations is negligible compared to that of the particle position dynamics, i.e. in the fast stiffness limit described above. 
The analysis of the high temperature case is particularly challenging: even for static harmonic potentials Kramers' theory has been shown to breakdown in this limit \cite{Abkenar2017}; the high temperature limit in the case of a fluctuating potential is an open problem. 

\begin{figure}
    \centering
    \includegraphics[width=0.9\textwidth]{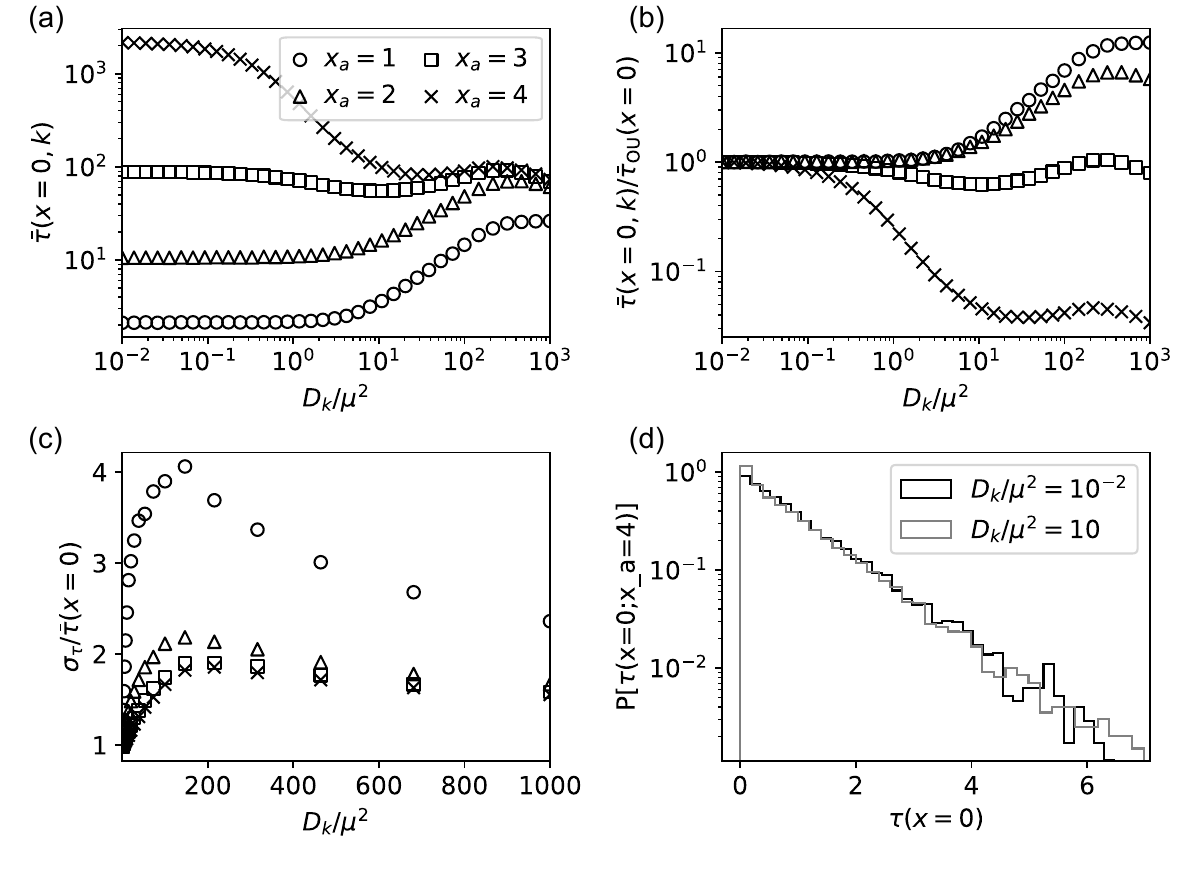}
    \caption{First-passage time statistics as a function of stiffness fluctuations strength. (a) Mean first-passage time as a function of fluctuation strength $D_k/\mu^2$ for various absorbing target locations $x_a>0$. (b) Mean first-passage time normalised by the mean first-passage time of the standard Ornstein-Uhlenbeck process (i.e. the $D_k/\mu^2 \to 0$ limit). We observe that stiffness fluctuations can significantly reduce the mean first-passage time to targets located far enough from the initial conditions (here, in $x_0 = 0$). (c) Coefficient of variation for the first-passage statistics showing non-monotonic behavior for all target locations. (d) Empirical histograms of the first passage time normalised by its standard deviation, showing an exponential decay of the first passage probability at long times. All results presented here are obtained via the integration of Eq.\,\eqref{eq:OU2} using an Euler-Maruyama method with $D_x= \bar{k} = 1$ over an ensemble of $m=10^5$ realisations of the coupled dynamics}
    \label{fig:mFPT}
\end{figure}

Here, we instead tackle this problem numerically. Specifically, we numerically integrate Eq.\,\eqref{eq:OU2} using the Euler-Maruyama method with timestep $dt = 10^{-4}$. For all results presented here (see Fig.\,\ref{fig:mFPT}), we simulate $m = 10^{5}$ realisations of the coupled dynamics in which the particle is initialised at $x_0 = 0$ and we place an absorbing boundary condition in $x_a > 0$, fixing $D_x = \bar{k} = 1$ with $\mu=0.1$. We have confirmed that all realisations lead to a finite first-passage time to the absorbing boundary condition.  We probe a wide range of values for the stiffness fluctuations strength, such that for high enough values of $D_k /\mu^2$, the steady-state distribution for the particle position may not exist. However, for all values of the stiffness fluctuations strength $D_k/\mu^2$ studied here, we have checked that the standard deviation of the mean-first passage time (mFPT) is finite and independent of the number of realisations for $m$ sufficiently large ensuring the convergence of the mFPT.

%%%%%%%%%%%%%%%%%
%%%%%%%%%%%%%%%%%

A first look at Fig.\,\ref{fig:mFPT} shows that the mFPT independently of the target location is generically a non-trivial non-monotonic function of the stiffness fluctuation strength, $D_k/\mu^2$. 
%Yet, we can extract from our results important results. 
In particular, we find that the mFPT to the absorbing target can be significantly reduced (more than one order of magnitude) by strong enough stiffness fluctuations $D_k/\mu^2$ for targets which are not too close to the initial conditions, which is consistent with previous studies. Interestingly, we show that instead stiffness fluctuations can increase the mFPT compared to the standard Ornstein-Uhlenbeck case (i.e. the zero stiffness fluctuations limit, $D_k/\mu^2 \to 0$) if $x_a$ is small. Said differently, for targets very close to the initial conditions, stiffness fluctuations can be detrimental and a system with constant stiffness $\bar{k}$ will instead be optimal. Furthermore, we observe that in all cases the coefficient of variation of the first-passage times, $\sigma_\tau/\bar{\tau}$, first strongly increases before reaching a maximum at the same value of $D_k / \mu^2$; at large enough fluctuations strength, the coefficient of variation decays monotonously with fluctuation strength for all target locations. Interestingly, we find that the first-passage times are exponentially distributed at both low and high values of $D_k/\mu^2$ for the most distant absorbing target location.

%%%%%%%%%%%%%%%%%
% Section: Entropy production.  %
%%%%%%%%%%%%%%%%%
\section{Entropy production}
\label{sec:epr}

\begin{figure}
    \centering
    \includegraphics[width=0.6\textwidth]{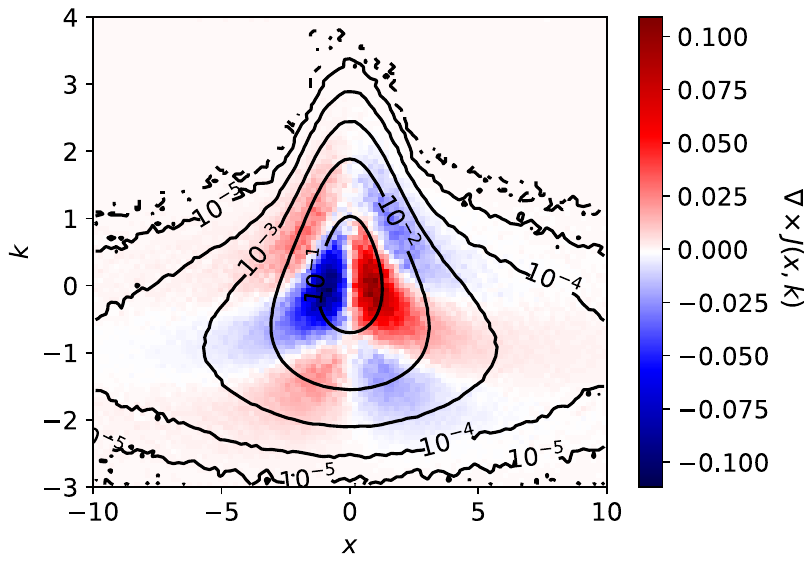}
    \caption{Steady state probability density (contours) and out-of-plane component of the curl of the steady-state probability current. The existence of steady state divergence-free currents implies time-reversal symmetry breaking and thus a non-zero entropy production, which we calculate analytically in Section \ref{sec:epr}.}
    \label{fig:epr}
\end{figure}

We now consider the thermodynamic implications of the stiffness fluctuations characterising the OU$^2$ process. Indeed, 
we expect a non-zero rate of entropy production at steady-state \cite{Cocconi2020,Alston2022,Cocconi2023} as the system performs work to change the stiffness.
%given the need for work to be done on the system to change the stiffness, a non-zero rate of entropy production is expected at steady-state \cite{Cocconi2020,Alston2022,Cocconi2023}. 
Furthermore, the existence of steady-state divergence-free probability currents in the $(x,k)$-plane, as shown on Fig.\,\ref{fig:epr}, implies a breaking of time-reversal symmetry. It is in this sense that we have previously referred to the OU$^2$ process as a minimal model of dissipative confinement. As detailed in Ref.~\cite{Alston2022} and \ref{sec:derivation_epr}, such dissipation can be written in terms of the second moment of the steady-state marginal probability density function for the coordinate $x$, which we computed in Sec.~\ref{sec:moments}. We now rederive this result using a shortcut and call upon the results of Sec.~\ref{sec:moments} to provide a clearer picture of the thermodynamic properties of the OU$^2$ process.

As a preliminary step, let us introduce the Fokker-Planck formulation \cite{Risken1996} of the Langevin dynamics \eqref{eq:OU2_x} and \eqref{eq:OU2_k},
\begin{equation}\label{eq:FP_ou2}
    \partial_t P(x,k) = [D_x \partial_x^2 + D_k \partial_k^2]P(x,k) + (\bar{k}+k) \partial_x [xP(x,k)] + \mu \partial_k [kP(x,k)] ~.
\end{equation}
Multiplying both sites of Eq.~\eqref{eq:FP_ou2} by $x^2$ and subsequently integrating with respect to both $x$ and $k$, using integration by parts where necessary, gives the remarkably simple relation 
\begin{equation}\label{eq:redl_for_epr}
    \left\langle (k(t) + \bar{k})x^2(t) \right\rangle=D_x~.
\end{equation}
Note however that, when carrying out this procedure, one encounters the following integral
\begin{align}
    \int dx dk  \ x^2 (k+\bar{k}) \partial_x [xP(x,k)] &= \bar{k} \int dx dk \ x^2 \partial_x [xP(x,k)] \nonumber \\
    &= \bar{k} \left[ x^3 \int dk P(x,k) \right]_{x=-\infty}^{+\infty} - 2 \bar{k} \int dx dk \ x^2 P(x,k)~.
\end{align}
For the boundary term on the right-hand side to vanish, it is required that the marginal probability density $P(x) \equiv \int dk P(x,k)$ decays faster than $P(x) \sim x^{-3}$ as $x \to \infty$. This is also the condition for the existence of the second moment, such that the validity of Eq.~\eqref{eq:redl_for_epr} is contingent upon the existence of $\langle x^2 \rangle$. 

The mean rate of entropy production, denoted $\dot{S}_i$, is related to the Jarzynski stochastic work \cite{Speck2011,Lee2023}
\begin{equation}\label{eq:jarz_work}
    W_\tau = \int_0^\tau dt \ \dot{x} \circ [(k(t) + \bar{k}) x(t)]= \frac{1}{2}\int_0^\tau dt \ \dot{k}(t) \circ x^2(t)
\end{equation}
via 
\begin{equation}\label{eq:ou2_epr}
    \dot{S}_i = \lim_{\tau \to \infty} \frac{1}{D_x} \langle W_\tau \rangle = - \frac{\mu}{2D_x} \langle k x^2 \rangle = \frac{\mu\bar{k}}{2D_x} \left(\langle x^2 \rangle - \frac{D_x}{\bar{k}} \right)
\end{equation} 
where $\circ$ denotes a Stratonovich product and we have used \eqref{eq:redl_for_epr} to replace $\langle kx^2\rangle$ in Eq.~\eqref{eq:ou2_epr}. Note that $\langle x^2 \rangle > D_x/\bar{k}$ for all $D_k >0$ \cite{Alston2022}, such that the second law of thermodynamics $\dot{S}_i \geq 0$ is always satisfied.
Now, using Eq.~\eqref{eq:mom2_sum_form} for the variance, we can write the steady-state entropy production rate as 
\begin{equation}
	\dot{S}_i = \frac{\mu}{2} \left( s e^{-\xi} \sum_{\ell=0}^\infty \frac{\xi^\ell}{\ell! (s-\xi+\ell)} - 1 \right)
	\label{eq:SSEPOUOU}
\end{equation}
where again we have defined $\xi = {4D_k}/{\mu^3}$ and $s = {2\bar{k}}/{\mu} $. Expanding to leading order in weak stiffness noise, $\dot{S}_i = {\mu \xi}/[{2s(1+s)}] + \mathcal{O}(\xi^2) = D_k/[\bar{k}(\mu + \bar{k})]+ \mathcal{O}(\xi^2)$, which vanishes at $D_k=0$.
Remarkably, the entropy production rate diverges together with $\langle x^2 \rangle$ as $D_k/\mu^2$ approaches $\bar{k}/2$ from below. 

%%%%%%%%%%%%%
%%%%%%%%%%%%%

%%%%%%%%%%%%%
% Section: Conclusion  %
%%%%%%%%%%%%%
\section{Conclusion}

We have examined a number of statistical, dynamic and thermodynamic properties of the OU$^2$ process: a generalised Ornstein-Uhlenbeck (OU) process obtained by allowing for the associated stiffness coefficient to undergo OU-like stochastic fluctuations around a positive mean. To the best of our knowledge, this is the first systematic exploration of such a model, which was originally introduced in two recent works by some of the authors on the thermodynamics of Brownian motion in fluctuating potentials \cite{Alston2022,Cocconi2023}. This process finds physical relevance in contexts where effective harmonic confining potentials are generated by a non-ideal contextual processes, e.g. when colloid are manipulated by optical tweezers with realistic laser power stability \cite{Bustamante2021,Pesce2020}. We also argued that the OU$^2$ model is closely related to models in stochastic search with resetting \cite{Evans2020, Alston2022,Gupta2020a,Gupta2020b,Jerez2021} and active matter \cite{Martin2021,Bothe2021,Cocconi2023}, and that it constitutes a stochastic counterpart to the breathing harmonic potentials studied in the thermodynamics literature \cite{Speck2011,Blickle2012,Martinez2016,Martinez2017}. 

We started out analysis by considering the limit of vanishing positional noise, for which the time-dependent probability density can be obtained in closed form, Eq.~\eqref{eq:sol_zeronoise}. From this, we derived exact expressions for the median (Eq.~\eqref{eq:medappreq}), mode (Eq.~\eqref{eq:mode}) and mean (Eq.~\eqref{eq:mean_x_nonoise}), whose time-dependence was found to transition from exponentially decreasing (trapped regime) to exponentially increasing (growing regime) at different non-trivial critical values of the non-dimensional parameter $\mathcal{F} \equiv D_k/(\mu^2 \bar{k})$. 

Having reintroduced a finite positional noise, we computed the conditional and full Green's function of the process, Eqs.~\eqref{eq:cond_prop} and \eqref{eq:full_green}, showing that both grow exponentially with time when $\mathcal{F}$ exceeds unity, indicating loss of ergodicity. Starting from the formal solution of the OU$^2$ dynamics, Eq.~\eqref{eq:formal_sol}, we subsequently computed the second and fourth moments of the steady-state positional probability density function, Eqs.~\eqref{eq:mom2_sum_form} and \eqref{eq:mom4_sumform}, as well as a necessary condition for the existence of moments of order $2n$ in terms of an n-dependent upper bound on $\mathcal{F}$, Eq.~\eqref{eq:mom_cov_cont_main}. From this condition we inferred an algebraic asymptotic decay of the positional probability density, with scaling exponent $\eta = -1-\mathcal{F}^{-1}$. We subsequently considered two limiting regimes of fast stiffness dynamics: while the na{\"i}ve adiabatic limit produced trivial OU dynamics for the slow degree of freedom, the second limit, obtained by taking $\mu \to \tilde{\mu}/\epsilon^2$ and $D_k \to \tilde{D}_k/\epsilon^4$ with $\epsilon \to 0$, led to non-trivial coarse grained dynamics for the position involving a renormalised stiffness and multiplicative noise, Eq.~\eqref{eq:fast_slow_lim_x}, for which the steady-state probability density was obtained in closed form, Eq.~\eqref{eq:analyt_fastslow_sol}. 

Borrowing from extreme value theory heuristics, we then conjectured that in the presence of finite stiffness fluctuations, $\mathcal{F}>0$, the standardised distribution of the running maximum should converge at long times to a Frechet form with $\mathcal{F}$-dependent exponent, in good agreement with numerical experiments (cf. the standard OU process, whose maximum is Gumbel distributed at long times). Further, the dependence on $\mathcal{F}$ of the mean first passage time to a positive target was studied numerically and we show that sufficiently strong stiffness fluctuations can aid the particle by reducing drastically the mean first-passage time to that target. A formal analytical treatment of this problem remains an intriguing open question and is left for future studies. 

Finally, we presented a compact derivation of the steady-state entropy production rate which calls upon results for Jarzinsky's stochastic work, Eq.~\eqref{eq:SSEPOUOU}, showing that it depends solely on the second steady-state moment of the positional probability density. Remarkably, the entropy production diverges for $\mathcal{F} > 1/2$. It would be interesting to explore higher order statistics of the stochastic work \eqref{eq:jarz_work}, similarly to what was done in Ref.~\cite{Manikandan2017} for the stochastically sliding potential and more recently in Ref.~\cite{Semeraro2023} for an AOU particle confined in a harmonic potential, and to compare any such result to the full distribution of the stochastic work obtained in Ref.~\cite{Speck2011} for a breathing harmonic potential in the slow driving regime.

Taken together, our results point to the OU$^2$ process as a widely applicable minimal model of dissipative confinement. It's rich phenomenology, emerging from the dynamical establishment of a non-equilibrium steady-state analogous to that of Brownian motion under stochastic resetting, renders it a valuable non-motile counterpart to other minimal models of single-particle out-of-equilibrium dynamics, such as the AOU particle, that have been explored extensively in recent years.

\vspace{1cm}

\bibliographystyle{iopart-num}
\bibliography{references}

\providecommand{\newblock}{}
\begin{thebibliography}{10}
\expandafter\ifx\csname url\endcsname\relax
  \def\url#1{{\tt #1}}\fi
\expandafter\ifx\csname urlprefix\endcsname\relax\def\urlprefix{URL }\fi
\providecommand{\eprint}[2][]{\url{#2}}

\bibitem{Uhlenbeck1930}
Uhlenbeck G~E and Ornstein L~S 1930 {\em Phys. Rev.\/} {\bf 36}(5) 823--841

\bibitem{Einstein1905}
Einstein A {\em et~al.\/} 1905 {\em Annalen der physik\/} {\bf 17} 208

\bibitem{LeBellac1991}
Le~Bellac M and Barton G 1991 {\em Quantum and statistical field theory\/}
  (Oxford: Clarendon Press)

\bibitem{Hohenberg1977}
Hohenberg P~C and Halperin B~I 1977 {\em Rev. Mod. Phys.\/} {\bf 49}(3)
  435--479

\bibitem{Martinez2017}
Mart{\'\i}nez I~A, Rold{\'a}n {\'E}, Dinis L and Rica R~A 2017 {\em Soft
  matter\/} {\bf 13} 22--36

\bibitem{Saha2022}
Saha T~K, Lucero J~N, Ehrich J, Sivak D~A and Bechhoefer J 2022 {\em Physical
  Review Letters\/} {\bf 129} 130601

\bibitem{Saha2021}
Saha T~K, Lucero J~N, Ehrich J, Sivak D~A and Bechhoefer J 2021 {\em
  Proceedings of the National Academy of Sciences\/} {\bf 118} e2023356118

\bibitem{Saha2023}
Saha T~K, Ehrich J, Gavrilov M~c~v, Still S, Sivak D~A and Bechhoefer J 2023
  {\em Phys. Rev. Lett.\/} {\bf 131}(5) 057101

\bibitem{TalFriedman2020}
Tal-Friedman O, Pal A, Sekhon A, Reuveni S and Roichman Y 2020 {\em The journal
  of physical chemistry letters\/} {\bf 11} 7350--7355

\bibitem{Ashkin1986}
Ashkin A, Dziedzic J~M, Bjorkholm J~E and Chu S 1986 {\em Optics letters\/}
  {\bf 11} 288--290

\bibitem{Dufresne2001}
Dufresne E~R, Spalding G~C, Dearing M~T, Sheets S~A and Grier D~G 2001 {\em
  Review of Scientific Instruments\/} {\bf 72} 1810--1816

\bibitem{Bustamante2021}
Bustamante C~J, Chemla Y~R, Liu S and Wang M~D 2021 {\em Nature Reviews Methods
  Primers\/} {\bf 1} 25

\bibitem{Pesce2020}
Pesce G, Jones P~H, Marag{\`o} O~M and Volpe G 2020 {\em The European Physical
  Journal Plus\/} {\bf 135} 1--38

\bibitem{Chiang1995}
Chiang R, Liu P and Okunev J 1995 {\em Journal of Banking \& Finance\/} {\bf
  19} 1327--1340

\bibitem{Bartoszek2017}
Bartoszek K, Gl{\'e}min S, Kaj I and Lascoux M 2017 {\em Journal of theoretical
  biology\/} {\bf 429} 35--45

\bibitem{Alston2022}
Alston H, Cocconi L and Bertrand T 2022 {\em Journal of Physics A: Mathematical
  and Theoretical\/} {\bf 55} 274004

\bibitem{Cocconi2023}
Cocconi L, Alston H and Bertrand T 2023 {\em Phys. Rev. Res.\/} {\bf 5}(4)
  043032

\bibitem{Cox1985}
Cox J~C, Ingersoll J~E and Ross S~A 1985 {\em Econometrica\/} {\bf 53} 385--407
  ISSN 00129682, 14680262

\bibitem{Manikandan2017}
Manikandan S~K and Krishnamurthy S 2017 {\em The European Physical Journal B\/}
  {\bf 90} 1--19

\bibitem{Lee2023}
Lee H~K, Kwon Y and Kwon C 2023 {\em Journal of the Korean Physical Society\/}
  {\bf 83} 331--336

\bibitem{Santra2021}
Santra I, Das S and Nath S~K 2021 {\em Journal of Physics A: Mathematical and
  Theoretical\/} {\bf 54} 334001

\bibitem{Zhang2017}
Zhang Z and Wang W 2017 {\em Communications in Statistics-Simulation and
  Computation\/} {\bf 46} 4783--4794

\bibitem{Evans2020}
Evans M~R, Majumdar S~N and Schehr G 2020 {\em Journal of Physics A:
  Mathematical and Theoretical\/} {\bf 53} 193001

\bibitem{Gupta2020a}
Gupta D, Plata C~A, Kundu A and Pal A 2020 {\em Journal of Physics A:
  Mathematical and Theoretical\/} {\bf 54} 025003

\bibitem{Gupta2020b}
Gupta D, Plata C~A and Pal A 2020 {\em Physical review letters\/} {\bf 124}
  110608

\bibitem{Jerez2021}
Jerez M~J~Y, Bonachita M~A and Confesor M~N~P 2021 {\em Physical Review E\/}
  {\bf 104} 044609

\bibitem{Guyon2004}
Guyon X, Iovleff S and Yao J~F 2004 {\em ESAIM: Probability and Statistics\/}
  {\bf 8} 25--35

\bibitem{Kesten1973}
Kesten H 1973 {\em Acta Mathematica\/} {\bf 131} 207--248

\bibitem{Morita2016}
Morita S 2016 {\em Europhysics Letters\/} {\bf 113} 40007

\bibitem{Morita2018}
Morita S 2018 {\em Journal of the Physical Society of Japan\/} {\bf 87} 053801

\bibitem{Burkhardt2000}
Burkhardt T~W 2000 {\em Journal of Physics A: Mathematical and General\/} {\bf
  33} L429

\bibitem{Majumdar2001}
Majumdar S~N and Bray A~J 2001 {\em Physical Review Letters\/} {\bf 86} 3700

\bibitem{Majumdar2010}
Majumdar S~N, Rosso A and Zoia A 2010 {\em Journal of Physics A: Mathematical
  and Theoretical\/} {\bf 43} 115001

\bibitem{Boutcheng2016}
Boutcheng H~J~O, Bouetou T~B, Burkhardt T~W, Rosso A, Zoia A and Crepin K~T
  2016 {\em Journal of Statistical Mechanics: Theory and Experiment\/} {\bf
  2016} 053213

\bibitem{Singh2020}
Singh P 2020 {\em Journal of Physics A: Mathematical and Theoretical\/} {\bf
  53} 405005

\bibitem{Chubynsky2014}
Chubynsky M~V and Slater G~W 2014 {\em Phys. Rev. Lett.\/} {\bf 113}(9) 098302

\bibitem{Banks2016}
Banks D~S, Tressler C, Peters R~D, H{\"o}fling F and Fradin C 2016 {\em Soft
  matter\/} {\bf 12} 4190--4203

\bibitem{Sposini2018}
Sposini V, Chechkin A and Metzler R 2018 {\em Journal of Physics A:
  Mathematical and Theoretical\/} {\bf 52} 04LT01

\bibitem{Blickle2012}
Blickle V and Bechinger C 2012 {\em Nature Physics\/} {\bf 8} 143--146

\bibitem{Martinez2016}
Mart{\'\i}nez I~A, Rold{\'a}n {\'E}, Dinis L, Petrov D, Parrondo J~M and Rica
  R~A 2016 {\em Nature physics\/} {\bf 12} 67--70

\bibitem{Speck2011}
Speck T 2011 {\em Journal of Physics A: Mathematical and Theoretical\/} {\bf
  44} 305001

\bibitem{Martin2021}
Martin D, O'Byrne J, Cates M~E, Fodor E, Nardini C, Tailleur J and van Wijland
  F 2021 {\em Phys. Rev. E\/} {\bf 103}(3) 032607

\bibitem{Bothe2021}
Bothe M and Pruessner G 2021 {\em Phys. Rev. E\/} {\bf 103}(6) 062105

\bibitem{Pavliotis2008}
Pavliotis G and Stuart A 2008 {\em Multiscale methods: averaging and
  homogenization\/} (Springer Science \& Business Media)

\bibitem{Arfken2011}
Arfken G~B, Weber H~J and Harris F~E 2011 {\em Mathematical methods for
  physicists: a comprehensive guide\/} (Academic press)

\bibitem{dePirey2023}
de~Pirey T~A, Cugliandolo L~F, Lecomte V and van Wijland F 2023 {\em Advances
  in Physics\/}  1--85

\bibitem{Rubin2014}
Rubin K~J, Pruessner G and Pavliotis G~A 2014 {\em Journal of Physics A:
  Mathematical and Theoretical\/} {\bf 47} 195001

\bibitem{Majumdar2020}
Majumdar S~N, Pal A and Schehr G 2020 {\em Physics Reports\/} {\bf 840} 1--32

\bibitem{Haan2006}
Haan L and Ferreira A 2006 {\em Extreme value theory: an introduction\/} vol~3
  (Springer)

\bibitem{Abkenar2017}
Abkenar M, Gray T~H and Zaccone A 2017 {\em Phys. Rev. E\/} {\bf 95}(4) 042413

\bibitem{Kramers1940}
Kramers H 1940 {\em Physica\/} {\bf 7} 284--304 ISSN 0031-8914

\bibitem{Hanggi1990}
H\"anggi P, Talkner P and Borkovec M 1990 {\em Rev. Mod. Phys.\/} {\bf 62}(2)
  251--341

\bibitem{Reimann1995}
Reimann P 1995 {\em Phys. Rev. E\/} {\bf 52}(2) 1579--1600

\bibitem{Reimann1995b}
Reimann P 1995 {\em Phys. Rev. Lett.\/} {\bf 74}(23) 4576--4579

\bibitem{Stein1990}
Stein D~L, Doering C~R, Plamer R~G, van Hemmen J~L and McLaughlin R~M 1990 {\em
  Journal of Physics A: Mathematical and General\/} {\bf 23} L203

\bibitem{Doering1992}
Doering C~R and Gadoua J~C 1992 {\em Phys. Rev. Lett.\/} {\bf 69}(16)
  2318--2321

\bibitem{Cocconi2020}
Cocconi L, Garcia-Millan R, Zhen Z, Buturca B and Pruessner G 2020 {\em
  Entropy\/} {\bf 22} 1252

\bibitem{Risken1996}
Risken H and Risken H 1996 {\em Fokker-planck equation\/} (Springer)

\bibitem{Semeraro2023}
Semeraro M, Gonnella G, Suma A and Zamparo M 2023 {\em Phys. Rev. Lett.\/} {\bf
  131}(15) 158302

\bibitem{Gardiner1985}
Gardiner C~W {\em et~al.\/} 1985 {\em Handbook of stochastic methods\/} vol~3
  (springer Berlin)

\bibitem{Abramowitz1988}
Abramowitz M, Stegun I~A and Romer R~H 1988 Handbook of mathematical functions
  with formulas, graphs, and mathematical tables

\end{thebibliography}

\clearpage
\appendix

%%%%%%%%%%%%%%%%%%%%%%%%%%%%%%%
% Section: Key results for the Ornstein-Uhlenbeck process   %
%%%%%%%%%%%%%%%%%%%%%%%%%%%%%%%
\section{Key results for the Ornstein-Uhlenbeck process}
\label{sec:keyOU}

In this appendix, we recall some useful key results for the original Ornstein-Uhlenbeck process as described by Eq.\,(\ref{eq:original_OU}). First, the transition probability $P(x, t|x', t')$ for the process takes the form
\begin{equation}
    P(x, t | x', t') = \sqrt{\frac{\bar{k}}{2 \pi D_x(1-e^{-2\bar{k} (t-t')})}} \exp\left[-\frac{\bar{k}(x-x'e^{-\bar{k}(t-t')})^2}{2D_x(1-e^{-2\bar{k}(t-t')})}\right],
\end{equation}
where $t'<t$. The steady-state probability density function is thus a Gaussian distribution with zero mean and variance, $\langle x^2\rangle = D_x/\bar{k}$ \cite{Gardiner1985}. The exponential decay at large $x$ of the steady-state distribution ensures that all even moments $\langle x^{2n}\rangle$ are finite for $n\in \mathbb{Z}_{\geq 0}$, with odd moments vanishing due to the $x\rightarrow -x$ symmetry. The associated cumulant generating function then takes the form 
\begin{equation}
   K(t) = \log \langle e^{a x}\rangle = \frac{a^2 D_x}{2\bar{k}}
\end{equation}
where cumulants of order 3 and above vanish. The solution $x(t)$ for a given realisation of the noise can be written as
\begin{equation}
    x(t) = \sqrt{2D_x} \int_{-\infty}^t dt' \:\zeta(t')\exp \left[-\bar{k}(t-t')\right],
\end{equation}
from which we read off the Green's function for the process $\mathcal{G}(t)$ of the form 
\begin{equation}
    \mathcal{G}(t)= \exp \left[-\bar{k}t\right]\Theta(t)
    \label{eq:OU_prop}
\end{equation}
where $\Theta(t)$ is the Heaviside function. 

 %%%%%%%%%%%%%%%%%%%%%%%%%%%%
% Section: Convergence criterion for higher moments  %
%%%%%%%%%%%%%%%%%%%%%%%%%%%%
 \section{Convergence criterion for higher moments}
 \label{sec:convergence}
 
In this section, we discuss the derivation of the convergence criterion for moments of order $n>2$. Looking at Eq.~\eqref{eq:expansion_cum_gen}, we argue that since the function is finite for all $(t_1,...,t_n)$, any divergence of the integral in the right-hand side of \eqref{eq:gen_mom_contracted} must be controlled by the behaviour of the integrand in the regime $t_i \to -\infty$. Keeping only leading order terms in this limit, we thus rewrite \eqref{eq:cum_gen_id} using \eqref{eq:expansion_cum_gen} as
\begin{align}
     \left\langle \exp\left[ - 2 \sum_{i=1}^{n} \int_{t_i}^t dt_i' \ k(t_i') \right] \right\rangle 
     &\simeq \exp\left[ \frac{4D_k}{\mu^2} \left(  \sum_{i=1}^n (t-t_i) + 2 \sum_{i<j} (t-t_j)\right) \right] \nonumber \\
     &= \exp\left[ \frac{4D_k}{\mu^2} \sum_{i=1}^n (1 + 2(i-1))(t-t_i)
 \right]
\end{align}
where $\simeq$ denotes approximate equality at $t_i \to -\infty$. Combining this with \eqref{eq:gen_mom_contracted}, we argue that the moment of order $2n$ converges when the following integral also converges
\begin{equation}\label{eq:conv_int_cond}
    \mathcal{I}_n = \int dt_{1<...<n} \exp \left[ - \sum_{i=1}^n \left( 2\bar{k} - \frac{4D_k}{\mu^2}(1+2(i-1)) \right) (t-t_i) \right]~,
\end{equation}
% namely when
% \begin{equation}
%     \frac{D_k}{\mu^2} < \frac{\bar{k}}{2n}~.
% \end{equation}
% To see this we perform the innermost integral in \eqref{eq:conv_int_cond}, which for generic $n$ gives
% \begin{equation}
%     \int_{t_{n-1}}^t dt_n \exp \left[ - \left( 2\bar{k} - \frac{4D_k}{\mu^2}(2n-1) \right) (t-t_n) \right] \propto 1 - \exp \left[ - \left( 2\bar{k} - \frac{4D_k}{\mu^2}(2n-1) \right) (t-t_{n-1}) \right]~.
% \end{equation}
% We see that the exponential now feeds into the second innermost integral, and so on. \textcolor{red}{This needs to be done in a cleaner way, maybe add some steps.} Eventually, at the level of the outermost integral, we will be faced with terms of the form
% \begin{equation}
%     \int_{-\infty}^t dt_1 \exp \left[ ... -2 \left( n\bar{k} - \frac{2D_k}{\mu^2} \underbrace{\sum_{m=1}^n (2m-1)}_{n^2} \right) (t-t_1)\right]~,
% \end{equation}
% which become divergent when $\bar{k} < 2nD_k/\mu^2$. \\
%
% \subsection{Re-writing previous section [HA]}
% \halt{
% We identify that for the convergence of the $n$-th moment, we require convergence of the following integral:
% \begin{equation}\label{eq:conv_int_cond_repeat}
%     \mathcal{I}_n = \int dt_{1<...<n} \exp \left[ - \sum_{i=1}^n \left( 2\bar{k} - \frac{4D_k}{\mu^2}(1+2(i-1)) \right) (t-t_i) \right]~,
% \end{equation}
where $\int dt_{1<...<n}\equiv \int_{-\infty}^t dt_1\dots \int_{t_{n-1}}^t dt_n$. We now define $a_i = 2\bar{k} - {4D_k}[1+2(i-1)]/{\mu^2}$ and re-write the integral \eqref{eq:conv_int_cond} as
\begin{equation}
     \mathcal{I}_n = \int dt_{1<...<n}\prod_{i=1}^n e^{-a_i(t-t_i)}~.
\end{equation}
We now call upon the following result for generic $m$ iteratively
\begin{equation}
    \int_{t_{m-1}}^t dt_m e^{-a(t-t_m)}= \frac{1-e^{-a(t-t_{m-1})}}{a}
\end{equation}
to evaluate $\mathcal{I}_n$ up to a multiplicative constant. We integrate over $dt_{n}$ and re-arrange to write
\begin{align}
    \mathcal{I}_n &\propto  \int dt_{1<...<(n-1)} \prod_{i=1}^{n-1}e^{-a_i(t-t_i)}\left[1-e^{-a_n(t-t_{n-1})}\right]\\
    &=\int dt_{1<...<(n-1)} \prod_{i=1}^{n-2}e^{-a_i(t-t_i)}\left[e^{-a_{n-1}(t-t_{n-1})}-e^{-(a_{n-1}+a_n)(t-t_{n-1})}\right].
\end{align}
Integrating over $t_{n-1}$ we subsequently derive
\begin{align}
    \mathcal{I}_n &\propto  \int dt_{1<...<(n-2)} \prod_{i=1}^{n-2}e^{-a_i(t-t_i)}[ &&\lambda_{n-1}\left(1-e^{-a_{n-1}(t-t_{n-2})}\right) - 1+ e^{-(a_{n-1}+a_n)(t-t_{n-2})}]\nonumber\\
    &\propto\int dt_{1<...<(n-2)} \prod_{i=1}^{n-3}e^{-a_i(t-t_i)}[&&(\lambda_{n-1}-1)e^{-a_{n-2}(t-t_{n-2})}-\lambda_{n-1}e^{-(a_{n-2}+a_{n-1})(t-t_{n-2})}\nonumber\\
    & \:&&-e^{-(a_{n-2}+a_{n-1}+a_n)(t-t_{n-2})}].
\end{align}
where $\lambda_{n+1}=(a_{n-1}+a_n)/a_{n-1}$.  

Treating the remaining time variables in the same manner, we conclude that the convergence of the integral $\mathcal{I}_n$ is determined by the positivity of the partial sums $A_m = \sum_{i=1}^m a_i$ for $m=1\dots n.$ More specifically, the most strict condition is always the positivity of the full sum, $m=n$, as the sequence $\{a_i\}$ is monotonically decreasing: $a_{i+1}<a_i$. The condition for the convergence of the $n$-th moment is thus
\begin{equation}\label{eq:mom_exist_cond}
    0<A_n=\sum_{i=1}^n\left[2\bar{k} - \frac{4D_k}{\mu^2}(1+2(i-1))\right]= 2 \bar k n - \frac{4 D_k}{\mu^2}n^2 \implies \frac{D_k}{\mu^2}<\frac{\bar k}{2n},
\end{equation}
in agreement with the results obtained in Section\,\ref{sec:moments} for the particular cases $n=1$ and $n=2$.

%%%%%%%%%%%%%%%%%%%%%%%%%%%%%%%%
% Section: Homogenisation Procedure for Coupled Dynamics %
%%%%%%%%%%%%%%%%%%%%%%%%%%%%%%%%
\section{Homogenisation procedure for coupled dynamics}
\label{sec:homogen_review}

In this appendix, we first review multiscale methods for the coarse graining of the coupled dynamics of two stochastic variables with multiple timescales and then apply it to the homogeneisation of the OU$^2$ process. 

%%%%%%%%%%%%%%%%%
% Sub-section: General theory  %
%%%%%%%%%%%%%%%%%
\subsection{General theory}
\label{sec:homogen_general_theory}

In this section, we follow the treatment of the homogenisation procedure presented in Chapter 11 of Ref.~\cite{Pavliotis2008}. We begin from the most general form for the coupled dynamics of two stochastic processes $x(t)$ and $k(t)$,
\begin{align}
    \dot{x} &= \frac{1}{\varepsilon} f_0(x,k) + f_1(x,k) + \alpha(x,k) \zeta_x(t) \label{eq:a1_x}\\
    \dot{k} &= \frac{1}{\varepsilon^2} g (x,k) + \frac{1}{\varepsilon} \beta(x,k) \zeta_k(t) \label{eq:a1_k}
\end{align}
where $\varepsilon > 0$ is a dimensionless factor that will be taken to zero to enforce a separation of timescales between the ``slow'' $x$ and ``fast'' $k$ dynamics. Note that both $\dot{x}$ and $\dot{k}$ involve fast contributions to the dynamics, but the dynamics for $k$ are an order of $\varepsilon$ faster than those of $x$. 
Combining Eqs.~\eqref{eq:a1_x} and \eqref{eq:a1_k}, we construct the backward Kolmogorov equation for $v(x, k, t)$, which takes the form
\begin{equation}
    \partial_t v = \mathcal{L}v =  \frac{1}{\varepsilon^2}\left(g \partial_k v + \frac{1}{2} B \partial_k^2 v\right) + \frac{1}{\varepsilon}\left(f_0 \partial_xv\right) + \left(f_1\partial_xv+ \frac{1}{2}A\partial_x^2 v\right), \label{eq:a1_backkolm}
\end{equation}
where we have defined the diffusivities $A(x,y) = \alpha(x,y) \bar{\alpha}(x,y)$ and $B(x,y) = \beta(x,y) \bar{\beta}(x,y)$, with $\bar{\bullet}$ denoting the conjugate transpose. We write the order $\mathcal{O}(\varepsilon^{-2})$ contribution to the backward Fokker-Planck operator acting on $v$ as
\begin{equation}
    \mathcal{L}_{-2} v \equiv g \partial_k v + \frac{1}{2}B\partial_k^2 v.
\end{equation}
Clearly, $\mathcal{L}_{-2}w(x)=0$ for any function $w(x)$ that does not depend on $k$. Additionally, we define $\rho^\infty(k;x)$ as the normalised measure obtained by solving the associated steady-state forward Kolmogorov equation for a fixed value of the slow variable $x$, $\mathcal{L}^*_{-2}\rho^\infty(k;x) = 0$.
%
% Naturally, the ergodicity assumption to make is that this operator has the one dimensional null space characterised by the vanishing of terms independent of $k$, along with the presence of an ergodic measure whose density $\rho^{\infty}(k;x)$ satisfies
% \begin{equation}
%     \mathcal{L}_{-2}^*\rho^{\infty}(k;x)=\partial_k^2\left(\frac{B}{2}\rho^{\infty}(k;x)\right) - \partial_k\left(g \partial_k\rho^{\infty}(k;x)\right)=0.
% \end{equation}
%
For the limit $\varepsilon \to 0$ of this problem to be well-posed, we require that $f_0(x, k)$ satisfies the so-called \textit{centering condition}, namely that its average with respect to $\rho^\infty$ vanishes
\begin{equation}
    \int dk f_0(x, k) \rho^{\infty}(k;x) = 0.    
\end{equation}
% Were this not the case, then a first-order perturbation scheme (where there are only two timescales at play in the dynamics, so-called \textit{averaging} in this context) would be sufficient for capturing a signature of the fast variable in the effective dynamics for the slow variable. The two methods are compared in Appendix B for the $OU^2$ model.

We now construct a perturbative solution to Eq.~\eqref{eq:a1_backkolm} of the form $v = v_0 + \varepsilon v_1+ \varepsilon^2 v_2 + \mathcal{O}(\varepsilon^3)$. %using the ergodicity assumptions concerning the null space of $\mathcal{L}_{-2}$. 
Matching terms of order $\mathcal{O}(\varepsilon^{-2})$, we conclude that $v_0(x)$ is independent of $k$. At $\mathcal{O}(\varepsilon^{-1})$ and relying on the centering condition on $f_0$ introduced above, we find an expression for $v_1(x, k)$ in terms of $v_0(x)$ and a function $\Phi(x,k)$ which solves
\begin{align}
    -\mathcal{L}_{-2}\Phi({x,k}) = f_0(x,k)
    \quad\text{   with   }\quad \int dk \ \Phi(x,k) \rho^\infty(k;x) =0.
\end{align}
Finally, at $\mathcal{O}(\varepsilon^0)$ we derive a closed equation for $\partial_t v_0$ from the ergodicity assumption that $\int dk\rho^{\infty} \mathcal{L}_{-2}v_2=0$, namely
\begin{equation}\label{eq:a1_reduced_x_kol}
    \partial_t v_0 = F(x)\partial_x v_0 + \frac{1}{2} A(x) A(x)^T \partial_x^2 v_0
\end{equation}
where we have defined the vector fields
\begin{equation}
    F(x) = \int dk \ \left( f_1(x,k) + (\partial_x \Phi(x, k)) f_0(x,k) \right) \rho^\infty(k;x)
\end{equation}
and
\begin{equation}
    A(x)A^T(x) = A_1(x) + \frac{1}{2} (A_0(x) + A_0^T(x))~,
\end{equation}
with
\begin{align}
    A_0(x) &= 2 \int dk \ f_0(x,k) \Phi(x, k) \rho^\infty(k;x) \\
    A_1(x) &= \int dk \ A(x,k) \rho^\infty(k;x)~.
\end{align}
The It{\^o} Langevin equation corresponding to the backward Kolmogorov equation \eqref{eq:a1_reduced_x_kol} for $v_0$ constitutes our slow variable dynamics in the regime $\varepsilon \ll 1$ and reads
\begin{equation}
    \dot{x}(t) = F(x(t)) + A(x(t))\eta_x(t)~.
\end{equation}
This is the key result that we draw on in Section \ref{sec:fastslow}, as detailed in the rest of this Appendix.

%%%%%%%%%%%%%%%%%%%%%%%%%%%%
% Sub-section: Homogenisation for OU$^2$ Process. %
%%%%%%%%%%%%%%%%%%%%%%%%%%%%
\subsection{Homogenisation for OU$^2$ Process}
\label{sec:homogen_OU2}

We now discuss the treatment of dynamics on multiple timescales in the specific context of the OU$^2$ process, Eqs.~\eqref{eq:OU2_x} and \eqref{eq:OU2_k}, which we restate for convenience here:
\begin{align}
    \dot{x} &= - \left(\bar{k} + k\right) x + \sqrt{2 D_x} \zeta_x  \\
    \dot{k} &= -\mu k + \sqrt{2D_k} \zeta_k.
\end{align}

\subsubsection{Adiabatic limit ---} First, following Sec.~\ref{sec:adiabatic}, we consider a na{\"i}ve separation of timescales between the two dynamics, akin to the adiabatic limit in thermodynamics, obtained by introducing a small dimensionless parameter $\varepsilon$ via the rescaling $\mu \to \tilde{\mu}/\varepsilon^2$ and $D_k \to \tilde{D}_k/\varepsilon^2$,
\begin{align}
    \dot{x} &= - \left(\bar{k} + k\right) x + \sqrt{2 D_x} \zeta_x\\
    \dot{k} &= -\frac{\tilde{\mu}}{\varepsilon^2} k + \sqrt{\frac{2\tilde{D}_k}{\varepsilon^2}} \zeta_k.
\end{align}
Comparing these equations with Eqs.~\eqref{eq:a1_x} and \eqref{eq:a1_k} and matching terms by their order in $\varepsilon$, we conclude that in this limit $f_0(x,k)=0$, $f_1(x,k) = - \left(\bar{k} + k\right) x$ and $\alpha(x,k) = \sqrt{2D_x}$, while $g(x,k) = -\tilde{\mu}k$ and $\beta(x,k)= \sqrt{2\tilde{D}_k}$. Employing the procedure outlined in the previous section, it is then straightforward to verify that $F(x) = -\bar{k}x$ and $A(x) = \sqrt{2D_x}$ leading to 
\begin{equation}
	\dot{x}(t) = - \bar{k} x(t) + \sqrt{2D_x}\zeta_x(t),
\end{equation}
In other words, no signature of the coupling between $x$ and $k$ survives in the effective dynamics for the slow variable $x$ in the limit $\varepsilon \rightarrow 0.$

\subsubsection{White noise limit ---}
For a non-trivial contribution to appear in the slow dynamics, we subsequently consider a second fast-slow regime, which we refer to as the white noise limit in Sec.~\ref{sec:whitenoise}. This time, we perform the rescaling $\mu \to \tilde{\mu}/\varepsilon^2$ and $D_k \to \tilde{D}_k/\varepsilon^4$, whereby
\begin{align}
	\dot{x} &= - \left(\bar{k} + k\right) x + \sqrt{2 D_x} \zeta_x~,\\
    \dot{k} &= -\frac{\tilde{\mu}}{\varepsilon^2} k + \sqrt{\frac{2\tilde{D}_k}{\varepsilon^4}} \zeta_k~.
\end{align}
Now, let $\chi \equiv \varepsilon k$, such that 
\begin{align}
    \dot{x} &= -\frac{1}{\varepsilon}\chi x - \bar{k}x + \sqrt{2 D_x} \zeta_x~,\\
    \dot{\chi} &= -\frac{1}{\varepsilon^2} \tilde{\mu} \chi + \frac{1}{\varepsilon}\sqrt{2\tilde{D}_k} \zeta_k~.
\end{align}
Comparing once again these equations with Eqs.~\eqref{eq:a1_x} and \eqref{eq:a1_k}, we identify $f_0(x,\chi) = - \chi x$, $f_1(x,\chi) = - \bar{k}x$ and $\alpha(x,\chi) = \sqrt{2D_x}$, while $g(x,\chi)= -\tilde{\mu}\chi$ and $\beta(x,\chi)= \sqrt{2\tilde{D}_k}$. 
Following the procedure outlined above, we find that the effective dynamics for the slow variable $x$ take the form
\begin{equation}\label{eq:xslow_konly}
    \dot{x}(t) = -\left( \bar{k} - \frac{D_k}{\mu^2} \right) x(t) + \sqrt{2\left( D_x + \frac{D_kx^2(t)}{\mu^2}\right)}\zeta_x(t)~,
\end{equation}
where we have used that ${\tilde{D}_k}/{\tilde{\mu}^2} = {D_k}/{\mu^2}$ in this case. We thus recover the effective slow dynamics \eqref{eq:fast_slow_lim_x} studied in the main text.

%%%%%%%%%%%%%%%%%%%%%%%%%%%%%%%
% Section: Connection to the Associated Legendre equation %
%%%%%%%%%%%%%%%%%%%%%%%%%%%%%%%
\section{Connection to the associated Legendre equation}
\label{sec:ass_legendre}

Upon enforcing the separation of timescales detailed in Sec.~\ref{sec:whitenoise}, the Fokker-Planck equation \eqref{eq:fp_fastslow_x} for the marginal distribution of the slow variable $x(t)$ can be mapped on to an associated Legendre differential equation through a change of variable, which we detail here. Indeed, the marginal steady-state distribution $P(x)$ is solution to
\begin{equation}\label{eq:FP_appB}
    0 = \frac{\partial}{\partial_{x}} \left\{ \frac{\partial}{\partial_{x}} \left[ \left( D_x + \frac{\tilde{D}_k x^2}{\tilde{\mu}^2} \right)P(x,t)\right] +  \left(\bar{k} - \frac{\tilde{D}_k}{\tilde{\mu}^2} \right) x P(x,t) \right\}~.
\end{equation}
Introducing $h \equiv \tilde{D}_k/\tilde{\mu}^2$, we define
\begin{equation}
    P(x) = a_0 \left( D_x + h x^2 \right)^{-\frac{1}{4}\left( 1 + \frac{\bar{k}}{h}\right)} \Phi(x)~,
\end{equation}
which we substitute into Eq.~\eqref{eq:FP_appB} to obtain
\begin{equation}
    0= \left( 1 + \frac{h}{D_x}x^2 \right) \Phi''(x) + \frac{2h}{D_x} x \Phi'(x) + \frac{h + \bar{k}}{4D_x} \left( 1 + \frac{D_x - \bar{k}x^2}{D_x + h x^2} \right) \Phi(x) .
\end{equation}
Finally, we perform the change of variable $s = i x \sqrt{h/D_x}$ and derive the equation for $\Phi(s)$ with imaginary argument
\begin{equation}
    0=(1-s^2) \Phi''(s) - 2 s \Phi'(x) + \left[ \left( \frac{\bar{k}}{2h} - \frac{1}{2}\right)\left( \frac{\bar{k}}{2h} - \frac{1}{2} + 1\right) - \left( \frac{h+\bar{k}}{2h}\right)^2 \frac{1}{1-s^2}\right]\Phi(s),
\end{equation}
which is now in the form of an associated Legendre differential equation \cite{Abramowitz1988}. The formal solution of the original equation, up to a normalisation factor, is thus 
\begin{equation}
    P(x) \propto \left( D_x + h x^2 \right)^{-\frac{\mu}{2}} \left[ c_1 P_\lambda^\mu\left( i\sqrt{\frac{h}{D_x}}x\right) + c_2 Q_\lambda^\mu\left( i\sqrt{\frac{h}{D_x}}x\right) \right]
\end{equation}
where 
\begin{equation}
    \mu = \frac{\bar{k}+h}{2}\quad\text{and}\quad \lambda = \frac{\bar{k}-h}{2},
\end{equation}
and $ P_\lambda^\mu$ and $ Q_\lambda^\mu$ denote the associated Legendre functions of the first and second kind, respectively, which are precisely the two linearly independent solutions of the Legendre equation. 

%%%%%%%%%%%%%%%%%%%%%%%%%%%%%%%%%%%%%%%
% Section: Derivation of the entropy production rate from Shannon entropy.  %
%%%%%%%%%%%%%%%%%%%%%%%%%%%%%%%%%%%%%%%
\section{Derivation of the entropy production rate from the Gibbs-Shannon entropy}
\label{sec:derivation_epr}

In this appendix, we summarise the full derivation of the entropy production rate $\dot{S}$ found in Ref.\,\cite{Alston2022}. We begin from the Fokker-Planck equation for the joint probability distribution for the particle position and stiffness governed by Eq.\,\eqref{eq:OU2}:
 \begin{equation}
     \partial_t P(x, k, t) = D_x \partial_x^2 P + (\bar k + k) \partial_x(xP)+D_k \partial_{k}^2P + \mu\partial_k(k P) = -\partial_x J(x, k, t) - \mathcal{J}(x, k, t)
 \end{equation}
 where we have defined the probability currents for the positional and stiffness variables respectively as
 \begin{equation}
     J(x, k, t) = -(\bar k + k) x P - D_x\partial_x P\quad \text{and}\quad \mathcal{J}(x, k, t)=-\mu k P - D_k \partial_k P.
 \end{equation}
 Following the standard approach for the thermodynamic treatment of diffusive systems with fluctuating potentials \cite{Alston2022}, we differentiate the Gibbs-Shannon entropy with respect to time to write
 \begin{equation}
     \dot{S}(t) = -\int dx \int dk \partial_tP(x, k, t)   \log\left(\frac{P(x, k, t)}{\bar P}\right), 
 \end{equation}
 where $\bar P$ is an arbitrary constant for dimensional consistency and we work in units such that $k_B=1.$ We identify two equal and opposite contributions to this rate that we write as 
 \begin{equation}
     \dot{S}(t) = \dot{S}_i(t) + \dot{S}_e(t)
 \end{equation}
 where have defined the internal (or total) entropy production rate 
 \begin{equation}
     \dot{S}_i(t) = \int dk \int dx \frac{1}{P(x, k, t)}\left[\frac{J^2(x, k, t)}{D_x} + \frac{\mathcal{J}^2(x, k, t)}{D_k}\right]
 \end{equation}
 and the external entropy production (or entropy flow)
 \begin{equation}
     \dot{S}_e(t) = \int dk \int dx \frac{(\bar k + k)xJ(x, k, t)}{D_x}+\frac{1}{D_k}\int dk\int dx \mu k\mathcal{J}(x, k, t).
 \end{equation}
 At steady-state, $\partial_tP=0$ and hence $\lim_{t\rightarrow \infty}\dot{S}(t)$ vanishes, which implies the steady-state relation $\lim_{t\rightarrow \infty}\dot{S}_i(t)=-\lim_{t\rightarrow \infty}\dot{S}_e(t)$. In general, the internal entropy production is the quantity of interest in the thermodynamic characterization of non-equilibrium stochastic processes. In what follows, we evaluate the integrals for the external entropy production at steady-state due to their simpler form, then employ the steady-state relation to evaluate the more classical thermodynamic quantity. 

 The dynamics for the stiffness are given by the equilibrium Ornstein-Uhlenbeck process, thus at steady-state the current $\mathcal{J}(k) = \int dx \mathcal{J}(x, k)$ vanishes. Thus the only contribution to $\dot{S}_e(t)$ at steady-state is the first term, that we can re-write as 
 \begin{equation}
     \lim_{t\rightarrow\infty}\dot{S}_e(t) = -\frac{\langle ((\bar k + k)x)^2\rangle}{D_x} + \bar k,
 \end{equation}
 where the average $\langle \cdot\rangle$ is taken over the joint probability distribution $P(x, k).$ Multiplying the Fokker-Planck equation by $x^2$, we integrate over $x$ to derive an equation for the dynamics of the marginal variance $\Xi(k,t)$:
 \begin{equation}
 \partial_t \Xi(k,t) = 2D_xP^{\rm tot}(k,t)  - 2(\bar k +k) \Xi(k,t)+\partial_k\big[D_k\partial_k\Xi(k,t) + \mu k\Xi(k,t)\big].
 \label{eq:TDVarA}
 \end{equation}
 Integrating this last equation at steady-state with respect to $k$ leads to $\langle k x^2\rangle = D_x$. 

 We then multiply (\ref{eq:TDVarA}) by $k$ before again integrating over $k$ to obtain
 \begin{equation}
 	\int dk \bigg[\frac{(\bar k+k)^2}{D_x}\Xi(k)\bigg] = \bar k +  \frac{1}{2D_x}\int d k \bigg[ D_k\partial_k\Xi(k) + k\partial_k\big[\mu k\Xi(k)\big]\bigg] .
 	\label{eq:int_step}
 \end{equation}
 Finally, we argue that the second term on the right-hand side of \eqref{eq:int_step} equation can be written as
 \begin{equation}
 \frac{1}{2D_x}\int d k \bigg[ D_k\partial_k\Xi(k) + k\partial_k\big[\mu\bar k\Xi(k)\big]\bigg]  =  -\frac{\mu}{2D_x}\big[\langle(\bar k + k) x^2\rangle - \bar k\langle x^2\rangle\big],
 \end{equation}
 noticing that the term proportional to $D_k$ vanishes by imposing a sufficiently fast decay of $\partial_x P$ at $x \to \pm \infty$. Using $\langle (\bar k + k) x^2 \rangle = D_x$, we conclude that the \textit{internal} entropy production rate at steady-state can be expressed as
 \begin{equation}
 	\lim_{t\rightarrow\infty}\dot{S}_i =\frac{\mu\bar k}{2D_x}\bigg(\langle x^2\rangle - \frac{D_x}{\bar k}\bigg).
 \end{equation}

\end{document}